\documentclass[12pt,a4paper]{article}
\usepackage{graphicx,multirow,outlines,ulem}
\usepackage[latin1]{inputenc}
\usepackage{amsmath,bm}
\usepackage{amsmath} 
\usepackage{amsfonts}
\usepackage{amssymb}
\usepackage{float}
\usepackage{cite}
\usepackage{epstopdf}
\DeclareGraphicsExtensions{.pdf,.png,.jpg}
\usepackage{breqn}
\usepackage{upgreek}
\makeatother
\makeatletter
\newcommand{\mathleft}{\@fleqntrue\@mathmargin0pt}
\newcommand{\mathcenter}{\@fleqnfalse}
\usepackage{xcolor}
\colorlet{linkequation}{red}
\usepackage[colorlinks]{hyperref}
\def\beq{\begin{equation}}
\def\eeq{\end{equation}}
\def\bea{\begin{eqnarray}}
\def\eea{\end{eqnarray}}

\begin{document}

\begin{center}
  {\Large \bf Quantum tunneling from family of Cantor potentials in fractional quantum mechanics}
\vspace{1.3cm}

{\sf Vibhav Narayan Singh\footnote[1]{e-mail address:\ \ vibhav.ecc123@gmail.com\hspace{0.05cm}},
Mohammad Umar\footnote[2]{e-mail address:\ \ pha212475@iitd.ac.in},
Mohammad Hasan\footnote[3]{e-mail address:\ \ mhasan@isro.gov.in,\hspace{0.05cm} mohammadhasan786@gmail.com},\\
Bhabani Prasad Mandal\footnote[4]{e-mail address:\ \ bhabani.mandal@gmail.com,\hspace{0.05cm} bhabani@bhu.ac.in}}

\bigskip

{\it $^{1,4}$ Department of Physics,
Banaras Hindu University, Varanasi-221005, INDIA. \\
$^{2}$ Indian Institute of Technology, Delhi-110016, INDIA \\ 
$^{3}$Indian Space Research Organisation,
Bangalore-560094, INDIA. \\ } 

\bigskip
	\noindent {\bf Abstract}
\end{center}
We explore the features of non-relativistic quantum tunneling in space fractional quantum mechanics through a family of Cantor potentials. We consider two types of potentials: general Cantor and general Smith-Volterra-Cantor potential. The Cantor potential is an example of fractal potential while the Smith-Volterra-Cantor potential doesn't belong to the category of a fractal system. The present study brings for the first time, the study of quantum tunneling through fractal potential in fractional quantum mechanics. We report several new features of scattering in the domain of space fractional quantum mechanics including the emergence of energy-band like features from these systems and extremely sharp transmission features. Further the scaling relation of the scattering amplitude with wave vector $k$ is presented analytically for both types of potentials.

\medskip
\vspace{1in}
\newpage

\section{Introduction}    
Over the last two decades, fractional dynamics have been a diverse area of research. The concept of fractional quantum mechanics was introduced by Laskin in the year $2000$ \cite{L1, L2}. The motivation behind this work was to extend the path integral (PI) formulation of quantum mechanics (QM) \cite{Feynman} to the more broader class of paths. In the PI formulation of QM, the  path integrals are taken over Brownian paths which lead to the Schrodinger equation of motion. However, the Brownian paths are the subset of a broader general class of paths known as Levy paths characterized by a Levy index $\alpha$. For $\alpha=2$, all Levy paths are Brownian paths.  When the PI formulation of QM is extended to Levy paths, one get the fractional Schrodinger equation \cite{L1, L2} and the associated quantum mechanics is known as space fractional quantum mechanics (SFQM). A time fractional Schrodinger equation was proposed by Naber \cite{naber2004time}. Later Wang and Xu \cite{wang2007generalized} combined the two kinds of fractional Schrodinger equation together to construct a space-time fractional Schrodinger equation. These generalization of QM may help to describe more extensive phenomena of the microscopic world.  
\paragraph{}
The Levy paths has fractal dimension $\alpha$. In the case of SFQM, the range of $\alpha$ is $1 < \alpha \leq 2$ \cite{L1, L2}. The domain of SFQM have grown fast over the last two decades and various applications are discussed by different authors. Some of the notable work are the energy band structure for the periodic potential \cite{dong2007some}, position-dependent mass fractional Schrodinger equation \cite{el2019some}, fractional quantum oscillator \cite{kirichenko2018confinement}, nuclear dynamics of the $H_{2}^{+}$ molecular ion \cite{medina2019nonadiabatic}, propagation dynamics of a light beam \cite{zhang2015propagation}, spatial soliton propagation \cite{ghalandari2019wave}, solitons in the fractional Schrodinger equation with parity-time-symmetric lattice potential \cite{yao2018solitons}, gap solitons \cite{xiao2018surface}, Rabi oscillations in a fractional Schrodinger equation \cite{zhang2017resonant},  self-focusing and wave collapse \cite{chen2018optical}, elliptic solitons \cite{wang2019elliptic}, light propagation in a honeycomb lattice \cite{zhang2017unveiling}, scattering features in non-Hermitian SFQM  \cite{hasan2018nh}, tunneling time  \cite{hasan2020tunneling, hasan2018tunneling} etc. Different methods are used in such studies such as domain decomposition method \cite{rida2008solution}, energy conservative difference scheme \cite{wang2015energy}, conservative finite element method \cite{li2018fast}, fractional Fan sub-equation method \cite{younis2017dark}, split-step Fourier spectral method \cite{kirkpatrick2016fractional}, transfer-matrix method \cite{tare} etc.
\paragraph{}
The term fractal was first coined by Mandelbrot \cite{mandelbrot}. Fractals are geometric objects which have self-similarity and homogeneity at all known scales. The geometric structures of fractals at a given scale or stage are obtained through a basic mathematical operation acting on the geometric object known as `initiator'. The process of mathematical operation is called `generator' which can be  repeated on multiple levels. Through `generator', a geometrical object with sub-units are created that resembles the structure of the entire object (the initiator) \cite{falconer2004fractal}. Due to the fact that the real numbers can be divided arbitrarily, the self-similarity of fractals hold at all scales. Since nature has many fractal structures, regular and irregular fragmented structures can be understood/approximated in the context of fractals \cite{mandelbrot, voss, hurd}. However, in nature, the self-similarity doesn't hold at all scales and in general, there exists an upper and lower limit within which the self-similarity applies.  
\paragraph{}
One-dimensional scattering by a Cantor fractal potential is one of the simplest scattering problems of quantum tunneling through  fractal system. This problem have been extensively studied in quantum mechanics by using the transfer matrix method to derive various scattering properties  \cite {cantor_f1, cantor_f2,cantor_f3,cantor_f4,cantor_f5,cantor_f6,cantor_f7,cantor_f7_1,cantor_f8,cantor_f9,Hasan_SPP}. The composition properties of the transfer matrix have been used to derive the scattering coefficients and associated properties. In Cantor fractal potential, scattering coefficients have been found to show scaling law and sharp features of resonances $k$ \cite {cantor_f1,cantor_f2,cantor_f6, cantor_f7, cantor_f7_1}. The tunneling amplitude from Cantor potential can also be derived by using the concept of super periodic potential (SPP) \cite{Hasan_SPP}. 
\paragraph{}
Despite the several advancement in the study of SFQM as well as quantum tunneling from fractal potentials, at present tunneling properties from fractal potentials in SFQM is not yet studied. It is expected that such studies will bring new features of scattering   properties in the domain of SFQM ($\alpha < 2$) which are absent in the case of standard QM ($\alpha=2$). In the present study we mainly focus on the simplest fractal system in one dimension, Cantor fractal along with an another member of Cantor family potential known as Smith-Volterra-Cantor (SVC) potential. The SVC potential is not a fractal potential while the Cantor potential is a fractal. In order to keep the study more general in nature, we consider the general Cantor (GC) and general SVC (GSVC) potential system. These are constructed in such a way that for a given initial length $L$ and height $V$ of the rectangular barrier potential, a fraction of $\frac{1}{3}$ from the middle is removed at every stage `$G$' from the remaining segments for standard Cantor-$3$ potential. For GC potential (or Cantor-$\rho$ potential), instead of $\frac{1}{3}$, a fraction $\frac{1}{\rho}$ is removed where $\rho> 1$ is a real positive number. Similarly in GSVC potential (or SVC-$\rho$) potential, a fraction of $\frac{1}{\rho^{G}}$  is removed from the middle at each stage $G$ instead of $\frac{1}{4^{G}}$ as in case of standard SVC-$4$ system. Again $\rho \in R^{+}$ and $\rho >1$. A simple  observation shows that SVC system doesn't satisfy the criteria for the same `self-similarity' at each stage $G$ and therefore is not a fractal system. 
\paragraph{}
In an earlier work, we have shown that Cantor-$3$ and SVC-$4$ potential system are the special case of SPP \cite{Hasan_SPP}. This is also true for Cantor-$\rho$ and SVC-$\rho$ system. SPP concept is the generalization of periodic potential having arbitrary number of internal periodicity \cite{Hasan_SPP}. As we have not yet extended the concept of SPP in the domain of SFQM, we use the fundamental principle to derive the expressions for transmission amplitude using transfer matrix approach for both types of potential. We report new features of scattering from these systems in the domain of SFQM. Notable features are emergence of energy band structures from these potentials which are absent in standard QM and extremely sharp transmission resonances. The scaling behavior with wave vector $k$ is also presented analytically.      
\paragraph{}
This paper is organized as follows. In section \ref{sfse}, an overview of SFQM is presented.  The transfer matrix in SFQM for a localized and repeated potential is discussed in detail in the section \ref{tm_SFQM}. In section \ref{sfp} and \ref{sec_symm_fractal}, we provide a brief review of the symmetric fractal potential of the cantor family and its repeated system. In next section \ref{T_ampl}, explicit expression of `$\zeta_{j}$' (argument of Chebyshev polynomial of second kind) is expressed in order to get transmission amplitude in SFQM for general SVC and general Cantor potential. Afterward, in section \ref{Tras_Featu}, we provide graphically a detailed analysis of the transmission features for both the fractal potential in the domain of SFQM. Finally, at last, in section \ref{R&D} results and discussion are mentioned.

\section{Space fractional Schrodinger equation}
\label{sfse}
When the path integral formulation of quantum mechanics is generalized over Levy flight paths, it results in space fractional quantum mechanics (SFQM). The governing equation for SFQM is the space fractional Schrodinger equation. The form of space fractional Schrodinger equation is given by \cite{L1}, 
\begin{equation}
i \hbar \frac{\partial \psi (x,t)}{\partial t}= H_{\alpha} (x,t) \psi(x,t),
\label{tdfse}
\end{equation}
Where, $H_{\alpha} (x,t)$ is the fractional Hamiltonian operator. The Hamiltonian is expressed through the use of Riesz fractional derivative $(-\hbar^{2} \Delta)^{\alpha/2}$  as, 
\begin{equation}
H_{\alpha} (x,t)=D_{\alpha} (-\hbar^{2} \Delta)^{\alpha/2} +V(x,t).
\label{halpha}
\end{equation}
Here `$\alpha$' is the Levy index and  $\Delta=\frac{\partial^{2}}{\partial x^{2}}$. In SFQM, the range of $\alpha$ is  $1 < \alpha \leq 2 $ \cite{L2}. $D_{\alpha}$ is a constant, also called as generalized diffusion coefficient and depends upon system characteristics. The Riesz fractional derivative of the wave function $\psi(x,t)$ is defined through the use of Fourier transform of $\psi(x,t)$ as,
\begin{equation}
(-\hbar^{2}\Delta)^{\alpha/2} \psi(x,t)=\frac{1}{2\pi \hbar} \int_{-\infty} ^{\infty} { \tilde{\psi}(p,t)\vert p \vert ^{\alpha} e^{ipx/\hbar}dp }. 
\label{Riesz_fractional_derivative}
\end{equation}
The Fourier transform of $\psi(x,t)$ is given by,
\begin{equation}
\tilde{\psi}(p,t)= \int_{-\infty} ^{\infty} \psi(x,t) e^{-i p x/\hbar} dx.
\end{equation}
and its inverse Fourier transform is,
\begin{equation}
\psi(x,t)=\frac{1}{2\pi \hbar} \int_{-\infty} ^{\infty} \tilde{\psi}(p,t) e^{i p x/\hbar} dp.
\end{equation}
For the case when potential $V(x,t)$ is time independent i.e., $V(x,t)=V(x)$, we have the time independent fractional Hamiltonian operator $H_{\alpha}(x)$ as, 
\begin{equation}
H_{\alpha}(x) = D_{\alpha} (-\hbar^{2} \Delta)^{\alpha/2} +V(x).
\label{halphax}
\end{equation}
The time-independent space-fractional Schrodinger equation is, 
\begin{equation}
D_{\alpha} (-\hbar^{2} \Delta)^{\frac{\alpha}{2}}\psi(x)+V(x)\psi(x)=E\psi(x) .
\label{tifse}
\end{equation} 
By using the concept of separation of variables, it can be shown that the time independent wave function $\psi(x)$ is related to $\psi(x,t)$ as $\psi(x,t)=\psi(x)e^{-iEt/\hbar}$ where $E$ is the energy of the particle. For a detail discussion on SFQM readers are referred to \cite{L3}. 

In the next section, we briefly discuss the transfer matrix formulation of the tunneling problem in SFQM.  

\section{Transfer matrix in SFQM}
\label{tm_SFQM}
\begin{figure}[htb]
\begin{center}
\includegraphics[scale=0.6]{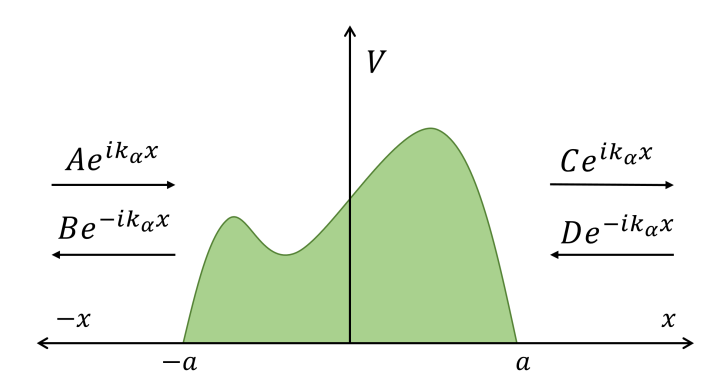}  
\caption{\small\textit{Depiction of the scattering of the quantum wave from an arbitrary potential $V(x)$ in one dimension.}}
\label{arb_pot}
\end{center}
\end{figure}
Consider a localized potential $V(x)$ bounded in the region $(-a,\,a)$ as shown in Fig. \ref{arb_pot}. The solution of time independent space fractional Schrodinger equation (Eq. \ref{tifse}) in all the three regions $x<-a$, $-a < x < a$, and $x > a$ are,
\begin{equation}
\varphi (x) = Ae^{ik_{\alpha}x} + Be^{-ik_{\alpha}x}, \,\,\,\,\,  x < -a,
\end{equation}
\begin{equation}
\vspace{0.1 cm}
\hspace{-1.0 cm}
\varphi (x) = \varphi_{ab} (x), \,\,\,\,\,-a < x < a,
\end{equation}
\begin{equation}
\varphi (x) = Ce^{ik_{\alpha}x} + De^{-ik_{\alpha}x}, \,\,\,\,\,x > a.
\end{equation}
Where,
\begin{equation}
k_{\alpha} = \left(\frac{E}{D_{\alpha}\hbar^{\alpha}}\right)^{1/\alpha}
\label{k1}
\end{equation}
and the coefficients $A$, $B$, $C$, and $D$ are the amplitudes of the waves on either side of the potential $V(x)$. The solution of the space fractional Schrodinger equation provides two linear equations in terms of the coefficients $A$, $B$, $C$, and $D$. The two linear equations can be represented in matrix form as,
\beq
\begin{pmatrix}   A(k_{\alpha}) \\ B(k_{\alpha})     \end{pmatrix}= M(k_{\alpha}) \begin{pmatrix}   C(k_{\alpha}) \\ D(k_{\alpha})    \end{pmatrix}.
\label{transfer_matrix}
\eeq
$M(k_{\alpha})$ is a $2 \times 2$ matrix,  
\beq
 M(k_{\alpha})= \begin{pmatrix}   M_{11}(k_{\alpha}) & M_{12}(k_{\alpha}) \\ M_{21}(k_{\alpha}) & M_{22}(k_{\alpha}) \end{pmatrix},
 \label{tmatrix_singlecell}
\eeq
which is known as the transfer matrix of the potential $V(x)$. For the case when $V(x)$ is Hermitian, the time invariance property of  Eq. \ref{tifse} leads to
\beq
M_{11}(k_{\alpha}) = M_{22}(k_{\alpha})^{*}, \ \  M_{21}(k_{\alpha}) = M_{12}(k_{\alpha})^{*},
\eeq
i.e., the diagonal and off-diagonal elements are complex conjugate to each other. The determinant of the transfer matrix is always unity which together with the above property implies $\vert M_{11}(k_{\alpha}) \vert ^{2} - \vert M_{12}(k_{\alpha}) \vert ^{2} = 1$. If the transfer matrix of a potential $V(x)$ is known then one can obtain the scattering coefficients for the potential $V(x)$ through the following expression, 
\beq
t_{l}(k_{\alpha}) = t_{r} (k_{\alpha}) = \frac{1}{M_{22}(k_{\alpha})}, \ \  r_{l}(k_{\alpha}) = -\frac{M_{21}(k_{\alpha})}{M_{22}(k_{\alpha})}, \ r_{r}(k_{\alpha}) = \frac{M_{12}(k_{\alpha})}{M_{22}(k_{\alpha})}. 
\eeq
From the knowledge of the transfer matrix of a single localized potential $V(x)$, one can obtain the transfer matrix of the periodic potential when $V(x)$ is periodically repeated $N_{1}$ times \cite{Griffiths_wave}. Formulation of the transfer matrix of locally periodic media from the knowledge of the transfer matrix of single `unit cell' potential $V(x)$ is also applicable in space fractional quantum mechanics \cite{tare}. The transfer matrix $M_{N_{1}}(k_{\alpha})$ for the periodic potential is given by,  
\beq
 M_{N_{1}}(k_{\alpha}) = \\ \begin{pmatrix}   [M_{11}e^{-ik_{\alpha}s}U_{N_{1}-1}(\zeta_{1})-U_{N_{1}-2}(\zeta_{1})]e^{ik_{\alpha}N_{1}s} & M_{12}U_{N_{1}-1}(\zeta_{1})e^{-ik_{\alpha}(N_{1}-1)s} \\ M_{12}^{*}U_{N_{1}-1}(\zeta_{1})e^{ik_{\alpha}(N_{1}-1)s} & [M_{11}^{*}e^{ik_{\alpha}s}U_{N_{1}-1}(\zeta_{1})-U_{N_{1}-2}(\zeta_{1})]e^{-ik_{\alpha}N_{1}s}   \end{pmatrix}.
 \label{tmatrix_periodic}
\eeq
In the above expression, `$s$' is the separation between the starting points of two consecutive `unit cell' potentials and $U_{N}(\zeta_{1})$ is the Chebyshev polynomial of the second kind. The argument of Chebyshev polynomial `$\zeta_{1}$', which is the Bloch phase of the corresponding fully developed periodic system, is computed from the knowledge of the `unit cell' transfer matrix and the separation `$s$' as \cite{Griffiths_wave},
\begin{equation}
\zeta_{1}(k_{\alpha}) = \frac{1}{2} \left ( M_{11} e^{-ik_{\alpha} s} + M_{22} e^{ik_{\alpha} s} \right ).
\label{zeta_exp}
\end{equation}
Using the property $M_{11}(k_{\alpha}) = M_{22}(k_{\alpha})^{*}$, the above equation can also be written as,
\begin{equation}
\zeta_{1}(k_{\alpha}) = \mbox{Re}[M_{22}]\cos(k_{\alpha}s) - \mbox{Im}[M_{22}]\sin(k_{\alpha}s) = \vert M_{22} \vert \cos(\phi+k_{\alpha}s),
\label{zeta_single_general}
\end{equation}
where $\phi$ is the argument of $M_{22}$, i.e., $M_{22} = \vert M_{22} \vert e^{i \phi} $.  The transmission coefficient for the periodic potential is the inverse of the lower diagonal element of the matrix given by  \ref{tmatrix_periodic}. Using the unitary properties of the transfer matrix, the transmission amplitude $T=\vert t_{l,r} \vert^{2}$ can be obtained as \cite{Griffiths_wave},  
\begin{equation}
T (N_{1})= \frac{1}{1 + [\vert M_{12} \vert U_{N_{1}-1}(\zeta_{1})]^{2}}.
\label{transmission_periodic}
\end{equation}
A few comments and the associated generalizations are in order. We can write the term $[\vert M_{12} \vert U_{N_{1}-1}(\zeta_{1})]^{2}$ appearing in the above equation as $[\vert M_{12} \vert U_{N_{1}-1}(\zeta_{1})]^{2}= \vert M_{12}  U_{N_{1}-1}(\zeta_{1}) \vert^{2}$ = $\vert (M_{12})_{N_{1}} \vert ^{2}$ where $(M_{12})_{N_{1}}$ is the $(1,2)$ element of the  transfer matrix (TM) given by  \ref{tmatrix_periodic}.  This can also be read as,
\begin{multline}
\vert  M_{12} \ \mbox {element of periodic system TM} \vert = \vert  M_{12}\  \mbox{element of unit cell TM} \times  \\ U_{N_{1}-1} (\mbox{Bloch phase of the fully developed periodic system}) \vert .
\label{m12_n_general}
\end{multline} 
If we periodically repeat this periodic system $N_{2}$ times with a different periodic distance $s_{2}$, then from Eq. \ref{m12_n_general}, the modulus of  $(1,2)$ element, $ \vert (M_{12})_{N_{1}, N_{2}} \vert $ of the transfer matrix of the new periodic system will be given by,
\begin{equation}
\vert (M_{12})_{N_{1}, N_{2}} \vert = \vert  (M_{12})_{N_{1}} U_{N_{2}-1}(\zeta_{2}) \vert =  \vert  M_{12} U_{N_{1}-1}(\zeta_{1}) U_{N_{2}-1}(\zeta_{2}) \vert . 
\label{m12_n1n2}
\end{equation} 
Where $\zeta_{2}$ is the Bloch phase for the new periodic system. If we periodically repeat the systems with parameters $N_{i}$ and  $s_{i}$ where  $i=1,2,3,...,G$ which yield $\zeta_{1}, \zeta_{2},.....\zeta_{G}$ as the respective Bloch phases, then Eq. \ref{m12_n1n2}  easily generalizes to
\begin{equation}
\vert (M_{12})_{N_{1}, N_{2}, N_{2},...,N_{G}} \vert =  \vert  M_{12} \prod_{i=1}^{G} U_{N_{i}-1}(\zeta_{i})  \vert . 
\label{m12_n1n2ng}
\end{equation} 
The corresponding transmission amplitude can be obtained from
\begin{equation}
T( N_{1}, N_{2}, N_{3},..., N_{G}) = \frac{1}{1+ \vert (M_{12})_{N_{1}, N_{2}, N_{2},...,N_{G}} \vert^{2} }= \frac{1}{ 1+ \vert  M_{12} \vert^{2} \prod_{i=1}^{G} U^{2}_{N_{i}-1}(\zeta_{i})  } . 
\label{t_n1n2ng}
\end{equation} 
A rigorous proof of Eq. \ref{t_n1n2ng}  based on the transfer matrix elements for super periodic potential is presented in \cite{Hasan_SPP} for the case of standard QM. In particular, when $N_{i}=2$, we have $U_{N_{i}-1}(\zeta_{i}) = U_{1} (\zeta_{i}) =2 \zeta_{i}$. Substitution of this in Eq. \ref{t_n1n2ng} leads to 
\begin{equation}
T(2, 2, 2,..., \mbox{$G$ times}) = T_{G}=  \frac{1}{ 1+ 4^{G} \vert  M_{12} \vert ^{2} \prod_{i=1}^{G} \zeta_{i}^{2} } ,
\label{tg_n1n2ng}
\end{equation} 
and the transfer matrix becomes,
\beq
 M_{N_{1}=2}(k_{\alpha}) = \begin{pmatrix}   2 M_{22}^{*} \zeta_{1} e^{ik_{\alpha}s}- e^{2 ik_{\alpha}s}  & 2 M_{12} \zeta_{1} e^{-ik_{\alpha} s} \\ 2 M_{12}^{*} \zeta_{1} e^{ik_{\alpha} s} & 2 M_{22} \zeta_{1} e^{-ik_{\alpha}s} - e^{-2 ik_{\alpha}s}   \end{pmatrix}.
 \label{tmatrix_2}
\eeq
It is to be noted that the form of Eq. \ref{tg_n1n2ng} is the general expression for tunneling amplitude when a single potential cell is repeated only two times and that system as a whole is further repeated two times and so on. We will extensively use Eq. \ref{tg_n1n2ng} and Eq. \ref{tmatrix_2} to calculate tunneling amplitudes for the symmetric potential of Cantor family. It turns out that tunneling amplitude for any symmetric potential which is generated by the division of a real line in three parts and subsequent removal of the middle segment can be expressed using Eq. \ref{tg_n1n2ng}. We will discuss this in detail in the subsequent sections.
\section{Symmetric potential of Cantor family}
\label{sfp}
\begin{figure}[htb]
    \centering
    \includegraphics[scale=0.45]{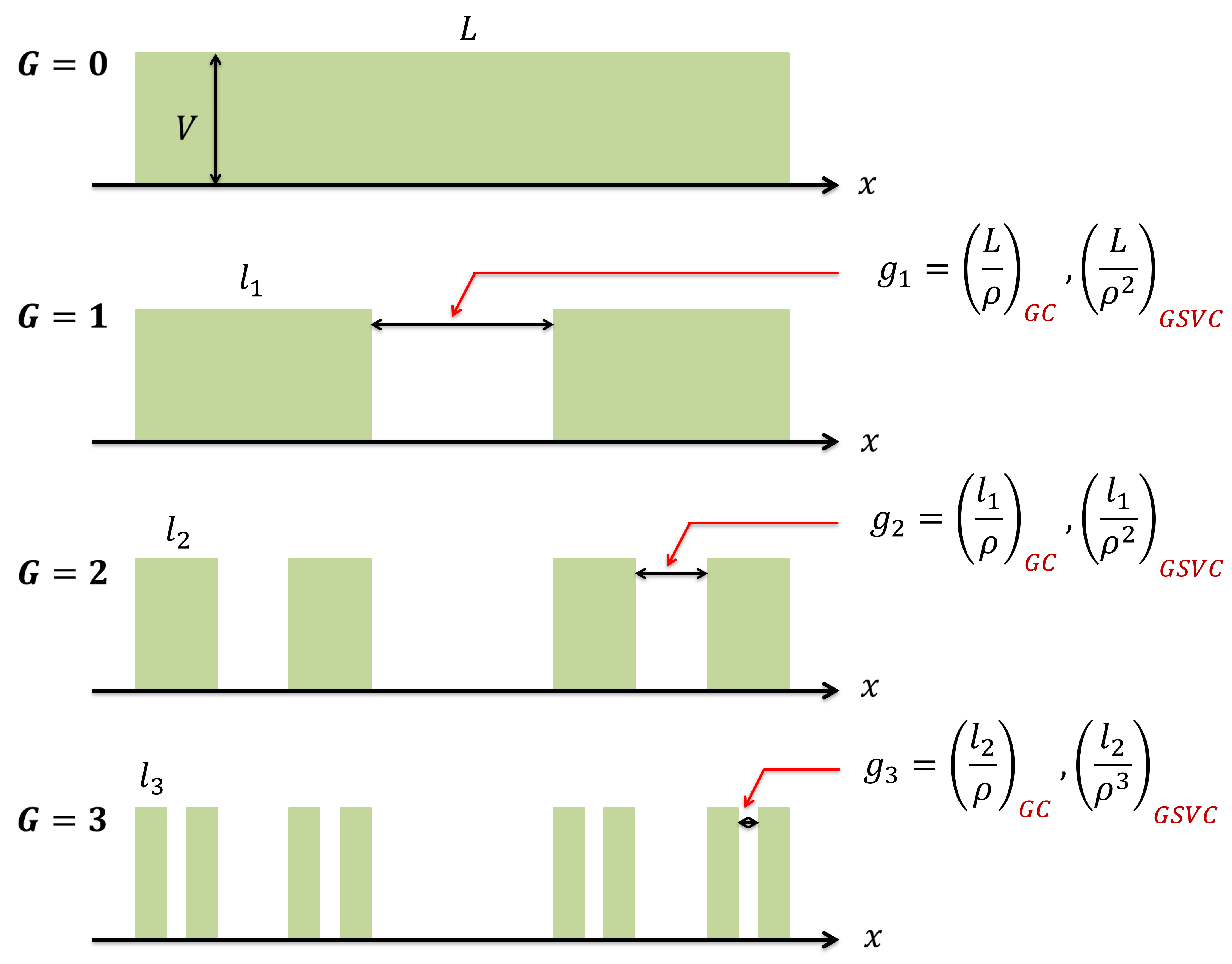}[H]
    \caption{\textit{ Construction of Cantor-$\rho$ and SVC-$\rho$ potential. The white region shows the gap between the potentials and the height of the opaque region is the potential height $V$. Here $G$ represents the stage of the system. In Cantor-$\rho$ potential, a fraction $1/\rho$ is removed at every stage while in SVC-$\rho$, a fraction $\frac{1}{\rho^{G}}$ is removed at each stage G.}}
    \label{symetric_fractal_system}
\end{figure}

In one dimension, a fractal  is generated by the division of a real line in a fashion which preserves self-similarity. Similarly, a rectangular fractal potential can be generated by dividing the length of the barrier in a self-similar fashion while keeping the height of the barrier unchanged. Symmetric fractal potential obeys parity symmetry about the origin i.e., the fractal potential is symmetric with respect to changing  $x \rightarrow -x$ and $-x \rightarrow x$. A potential of Cantor family is generated when the line segments are divided into three parts and the middle parts are removed at any stage $G$. A particular case of the symmetric Cantor potential is when the removal of the middle part from the line segment leaves the resultant two segments of equal sizes. This configuration of the system  is always symmetric. Starting from a length $L$ and stage $G=0$, symmetric Cantor potential can be generated by the removal of a fraction $\frac{1}{\rho^{a_{1}+a_{2}G}}$ from the middle segment(s) at each stage $G$. Here $\rho \in R^{+}$ and   $ a_{1}$, $a_{2}$  $\in \{0, R^{+} \}$. When $a_{1}=1$ and $a_{2}=0$, we have general Cantor potential  (also, for $a_{2}=0$, $a_{1}$ can be absorbed by defining $\rho^{a_{1}} = \rho_{1}$ for some real $\rho_{1}$ and we still have general Cantor potential). Similarly, for $a_{1}=0$, we have general Smith-Volterra-Cantor (SVC)  potential. For the special case when $a_{1}=0$, $a_{2}=1$ and $\rho=4$ we have  standard SVC potential system. Again when $a_{1}=0$, we can absorb $a_{2}$ by defining an associated new $\rho$ and the fractal potential is named an SVC-$\rho$ system.  The geometrical construction of general Cantor and general SVC potential is illustrated in Fig. \ref{symetric_fractal_system}. At any stage $G$, both general Cantor and general SVC potential have $2^{G}$ segments of equal length $l_{G}$. The value of $l_{G}$ are different for both types of potential.  In the case of general Cantor,    
\begin{equation}
l_{G}= \left(\frac{\rho - 1}{2\rho}\right)^{G}L.
\label{l_G_C}
\end{equation}
For the case of general SVC, $l_{G}$ can be obtained through the use of the $q$-Pochhammer symbol as shown below.
From Fig. \ref{symetric_fractal_system}, it is noted that,
\begin{equation}
l_{1} = \frac{L}{2}\left(1-\frac{1}{\rho}\right).
\end{equation}
Similarly, the segment length $l_{2}$ for stage $G=2$ is,
\begin{equation}
l_{2} = \frac{l_{1}}{2}\left(1-\frac{1}{\rho^{2}}\right) = \frac{L}{2^{2}}\left(1-\frac{1}{\rho}\right)\left(1-\frac{1}{\rho^{2}}\right).
\end{equation}
Similarly,
\begin{equation}
l_{3} = \frac{l_{2}}{2}\left(1-\frac{1}{\rho^{3}}\right) = \frac{L}{2^{3}}\left(1-\frac{1}{\rho}\right)\left(1-\frac{1}{\rho^{2}}\right)\left(1-\frac{1}{\rho^{3}}\right).
\end{equation}
By continuing the same steps, the segment length `$l_{G}$' for arbitrary $G^{th}$ order SVC-$\rho$ 0 potential is obtained as,
\begin{equation}
l_{G} = \frac{L}{2^{G}}\prod_{i=1}^{G}\left(1-\frac{1}{\rho^{i}}\right). 
\label{l_G_SVC}
\end{equation}
The product series can be recognized as, 
\begin{equation}
\prod_{i=1}^{G}\left(1-\frac{1}{\rho^{i}}\right) = q \left ( \frac{1}{\rho}; \frac{1}{\rho} \right)_{G}.
\end{equation}
Where,
\begin{equation}
q(a;\lambda)_{n}=\prod_{i=0}^{n-1}(1-a.\lambda^{i})=(1-a)(1-a.\lambda)(1- a. \lambda^{2}).....(1-a.\lambda^{n-1})
\label{qp}
\end{equation} 
is $q$-Pochhammer symbol \cite{abramowitz1964handbook}. Therefore, through the use of the $q$-Pochhammer symbol, we can express $l_{G}$ as,
\begin{equation}
l_{G} = \frac{L}{2^{G}} q \left ( \frac{1}{\rho}; \frac{1}{\rho} \right)_{G} .
\label{l_G_qp}
\end{equation}
\section{Symmetric Cantor family potentials as repeating systems}
\label{sec_symm_fractal}
\begin{figure}[H]
\begin{center}
\includegraphics[scale=0.5]{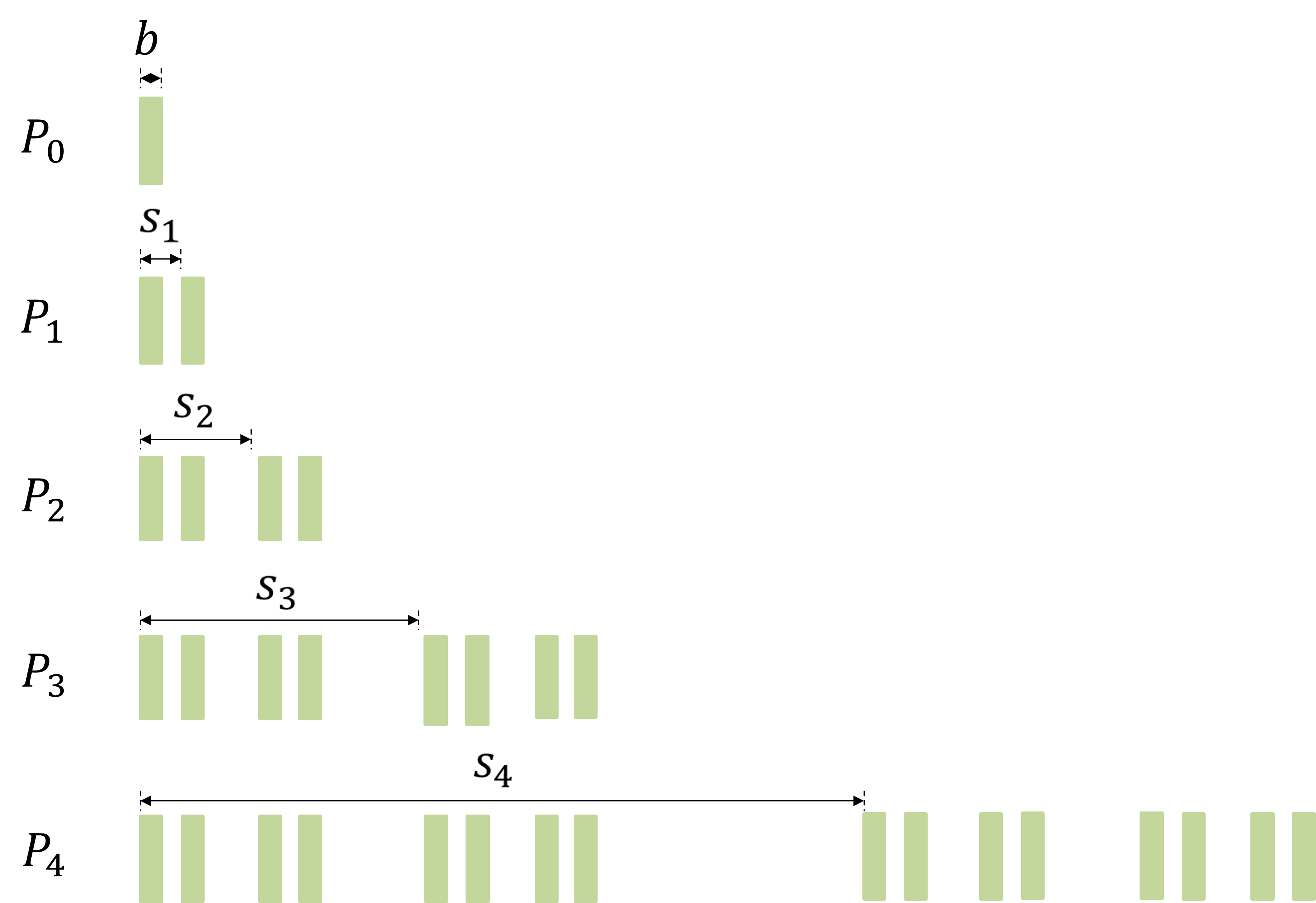}[H] 
\caption{\textit{Construction of the symmetric Cantor family potential for the stage $G = 4$ as periodic repetition of the periodic system of order 4.}}
\label{cons_sfp}
\end{center}
\end{figure}

In this section, we illustrate that a symmetric  potential of the Cantor family can be generated through a `unit cell' by repeating it two times and then repeating the resultant `cell' further two times and so on. Consider a rectangular barrier of height $V$ and width $l_{G}$ as shown in Fig. \ref{cons_sfp}. We can repeat this barrier at a distance $s_{1} > l_{G}$ as shown in Fig. \ref{cons_sfp}. The resultant system of these two barriers are further repeated at a distance of $s_{2}$ thereby generating a system of four rectangular barriers which as a whole is further repeated at a distance of $s_{3}$ as shown in Fig. \ref{cons_sfp}. This process of repeating the resultant barrier systems two times at a specific distance can continue up-to an arbitrary stage $G$. As the Cantor family systems are well defined mathematical structures, the value of `$l_{G}$'  and various `$s_{i}$' can be easily identified for any arbitrary stage $G$ for a particular system.   
\paragraph{}
First we present general expression of $s_{j}$ for Cantor-$\rho$ fractal system. For this system we have,
\begin{equation}
s_{1} = l_{G} + \frac{l_{G-1}}{\rho}, \nonumber 
\end{equation}
\begin{equation}
s_{2} = l_{G-1} + \frac{l_{G-2}}{\rho}, \nonumber
\end{equation}
\begin{equation}
s_{3} = l_{G-2} + \frac{l_{G-3}}{\rho}, \nonumber
\end{equation}
The above sequences show that, 
\begin{equation}
s_{j} = l_{G+1-j} + \frac{l_{G-j}}{\rho}.
\end{equation}
Using Eq. \ref{l_G_C}, this can be simplified to,
\begin{equation}
s_{j} = x^{G-j}y L,
\label{sj_cantor}
\end{equation}
where,
\begin{equation}
x = \frac{\rho - 1}{2\rho}, y = \frac{\rho + 1}{2\rho}.
\end{equation}
Similarly, it can be shown that for SVC-$\rho$  potential, $s_{j}$ is given by, 
	\begin{equation}
		s_{j} =  l_{G+1-p}+\frac{l_{G-p}}{\rho^{G+1- p}}.
	\end{equation}
	Using Eq. \ref{l_G_SVC} in the above expression, we have after simplification
	\begin{equation}
		s_{j}=\frac{L}{2^{G+1-j}}\left(1+\frac{1}{\rho^{G+1-j}}\right)   q \left ( \frac{1}{\rho}; \frac{1}{\rho} \right)_{G-j}.
	\label{sj_svc}
	\end{equation}	
For a given $G$, by choosing a single barrier of length $l_{G}$ as given by Eq. \ref{l_G_C} and placing the barrier at various $s_{j}$ as given by Eq. \ref{sj_cantor},  we get Cantor-$\rho$  potential. Similarly, by choosing $l_{G}$ from Eq. \ref{l_G_qp} and $s_{j}$ from Eq. \ref{sj_svc} we get SVC-$\rho$ potential. In the next section, we calculate the transmission amplitudes from these two types of fractal potentials in SFQM.
\section{Transmission amplitudes in SFQM}
\label{T_ampl}
It is clear from the discussion in the previous section that  (symmetric) Cantor-$\rho$ (GC) and SVC-$\rho$ (GSVC) potentials are the special cases of systems that are repeated two times and that configuration as a whole is further repeated two times and so on. The number of such operations of repetitions is equal to the stage $G$ of the GC and GSVC  potential. The general expression of the tunneling amplitude for such a potential system in SFQM is given by Eq. \ref{tg_n1n2ng}. What remains is to calculate the general expressions for $\zeta_{i}$, $i=1,2,3,.., G$ for GC and GSVC potentials. We will derive the general expression for $\zeta_{i}$ and then would specialize to calculate specific expressions for $\zeta_{i}$ for GC and GSVC  potentials. The calculations are illustrated below.
\paragraph{}
Let $M_{22}(k_{\alpha})$ denotes the lower diagonal elements of the transfer matrix of rectangular barrier of width $b=l_{G}$ and height $V$. This potential configuration is represented by $P_{0}$ in the Fig. \ref{cons_sfp}. Similarly, let $(M_{22})_{1}$, $(M_{22})_{2}$, $(M_{22})_{3}$ etc. denote the lower diagonal elements of the transfer matrix of the combined system represented by $P_{1}$, $P_{2}$, $P_{3}$ etc. as shown in Fig. \ref{cons_sfp}. The corresponding Bloch phases are $\zeta_{1}$, $\zeta_{2}$, $\zeta_{3}$ etc. respectively. From Eq. \ref{tmatrix_2} we can read that
\begin{equation}
(M_{22})_{j}= 2 (M_{22})_{j-1} \zeta_{j} e^{-ik_{\alpha}s_{j}} - e^{-2 ik_{\alpha}s_{j}},
\label{m22_j} 
\end{equation}
where $j=1,2,3,..,G$ and  $(M_{22})_{0}= M_{22}$. Now from the general Eq. \ref{zeta_single_general} we can write, 
\begin{equation}
\zeta_{2}(k_{\alpha})= \mbox{Re}\big[(M_{22})_{1}\big]\cos{k_{\alpha}s_{2}}- \mbox{Im}\big[(M_{22})_{1}\big]\sin{k_{\alpha}s_{2}}.
\label{x_2}
\end{equation}      
We can use Eq. \ref{m22_j} in the above equation so that, 
\begin{multline}
\zeta_{2}(k_{\alpha})= \mbox{Re}\big[(2\times M_{22}.\zeta_{1}) e^{-ik_{\alpha}s_{1}}-e^{- 2 . i k_{\alpha}s_{1}}\big]\cos{k_{\alpha}s_{2}}\\-\mbox{Im}\big[(2\times M_{22}\zeta_{1}) e^{-ik_{\alpha}s_{1}}-e^{- 2 . i k_{\alpha}s_{1}}\big]\sin{k_{\alpha}s_{2}}.
\label{x_2_N}
\end{multline}
The simplification of the real and imaginary parts finally gives,
\begin{equation}
\zeta_{2}=2 \lvert{M_{22}}\rvert \zeta_{1} \cos{\big[\phi-k_{\alpha}\{s_{1}-s_{2}\}\big]}-\cos{\big[k_{\alpha}\{2s_{1}-s_{2}\}\big]}.
\label{x_2_f}
\end{equation}
Similarly, repeating the above procedure to calculate $\zeta_{3}$ we have,  
\begin{equation}
\zeta_{3}=\mbox{Re}\big[(M_{22})_{2}\big]\cos{k_{\alpha}s_{3}}- \mbox{Im}\big[(M_{22})_{2}\big]\sin{k_{\alpha}s_{3}}.
\label{x_3}
\end{equation}
Again using Eq. \ref{m22_j} to simplify the above, we obtain for $\zeta_{3}$, 
\begin{multline}
\zeta_{3}(k_{\alpha})= 2^{2} \lvert{M_{22}}\rvert \zeta_{1}\zeta_{2}\cos{\big[\phi-k_{\alpha}\{s_{1}+s_{2}-s_{3}\}\big]}- \\ 2.\zeta_{2}\cos{\big[k_{\alpha}\{2s_{1}+s_{2}-s_{3}\}\big]}-\cos{\big[k_{\alpha}\{ 2s_{2}-s_{3}\}\big]} .
\label{x_3_f}
\end{multline}
Similarly, we have for $\zeta_{4}$
\begin{equation}
\zeta_{4}(k_{\alpha})=\mbox{Re}\big[(M_{22})_{3}\big]\cos{k_{\alpha}s_{4}}-\mbox{Im}\big[(M_{22})_{3}\big]\sin{k_{\alpha}s_{4}} .
\label{x_4}
\end{equation}
The repeated application of Eq. \ref{m22_j} and simplifications of the real and imaginary parts in the above equation gives,
\begin{multline}
\zeta_{4}(k_{\alpha}) = 2^{3} \lvert{M_{22}}\rvert\zeta_{1}\zeta_{2}\zeta_{3}\cos{\big[\phi- k_{\alpha}\{s_{1}+s_{2}+s_{3}-s_{4}\}\big]}\\-2^{2}\zeta_{2}\zeta_{3}\cos{\big[ k_{\alpha}\{2s_{1}+s_{2}+s_{3}-s_{4}\}]}-2. \zeta_{3}\cos{\big[k_{\alpha}\{2s_{2}+s_{3}-s_{4}\}\big]}\\-\cos{\big[k_{\alpha}\{2 s_{3}-s_{4}\}\big]} .
    \label{x_4_f}
\end{multline}
Similarly, the expression for $\zeta_{5}$ is given by,
\begin{multline}
\zeta_{5}(k_{\alpha}) = 2^{4} \lvert{M_{22}}\rvert\zeta_{1}\zeta_{2}\zeta_{3}\zeta_{4}\cos{\big[\phi- k_{\alpha}\{s_{1}+s_{2}+s_{3}+s_{4}-s_{5}\}\big]}\\-2^{3}\zeta_{2}\zeta_{3}\zeta_{4}\cos{\big[ k_{\alpha}\{2s_{1}+s_{2}+s_{3}+s_{4}-s_{5}\}]}\\-2^{2} \zeta_{3}\zeta_{4}\cos{\big[ k_{\alpha}\{2s_{2}+s_{3}+s_{4}-s_{5}\}\big]}-2.\zeta_{4}\cos{\big[k_{\alpha}\{ 2 s_{3}+s_{4}-s_{5}\}\big]}\\-\cos{\big[k_{\alpha}\{ 2 s_{4}-s_{5}\}\big]}.
    \label{x_5_f}
\end{multline}
We observe from the sequence of Eqs. \ref{x_2_f}, \ref{x_3_f}, \ref{x_4_f}, and \ref{x_5_f} that the general expression for $\zeta_{j}$ can be written in the following series form,
\begin{equation}
\zeta_{j}(k_{\alpha}) = 2^{j-1}\vert M_{22} \vert\cos\big[\phi-k_{\alpha} \eta_{1}(j)\big] \prod_{p=1}^{j-1} \zeta_{p} - \sum_{r=1}^{j-1}\left[2^{j-r-1}\cos \big[k_{\alpha} \eta_{2}(j,r)\big] \prod_{p=r+1}^{j-1}\zeta_{p}\right] .
\label{zeta_j1}
\end{equation}
In the above equation, we have used the following notation,
\begin{equation}
\eta_{1}(j) \equiv \left(\sum_{p = 1}^{j-1}s_{p}\right)-s_{j},
\end{equation}
\begin{equation}
\eta_{2}(j,r)  \equiv  \left(\sum_{p = r}^{j}s_{p}\right)- ( 2s_{j}-s_{r}). 
\end{equation}
It is easy to show that,
\begin{equation}
\eta_{2}(j,r) \equiv \eta_{1}(j)- \eta_{1}(r).  
\label{eta2_jr}
\end{equation}
Eq. \ref{zeta_j1} is the general expression for $\zeta_{j}$, $j=1,2,3,..,G$. However, it is important to note here that in Eq. \ref{zeta_j1}, we have to drop the terms when the running variable `$r$' is more than the upper limit for the summation operation and we take terms as unity when the running variable is more than the upper limit for the product operation. From the knowledge of $\zeta_{1}, \zeta_{2}, \zeta_{3},..., \zeta_{G}$, we can calculate the tunneling amplitude from Eq.  \ref{tg_n1n2ng}. Now we calculate the values of $\eta_{1,2}$ and their properties. 
\begin{figure}[htb]
\begin{center}
\includegraphics[scale=0.55]{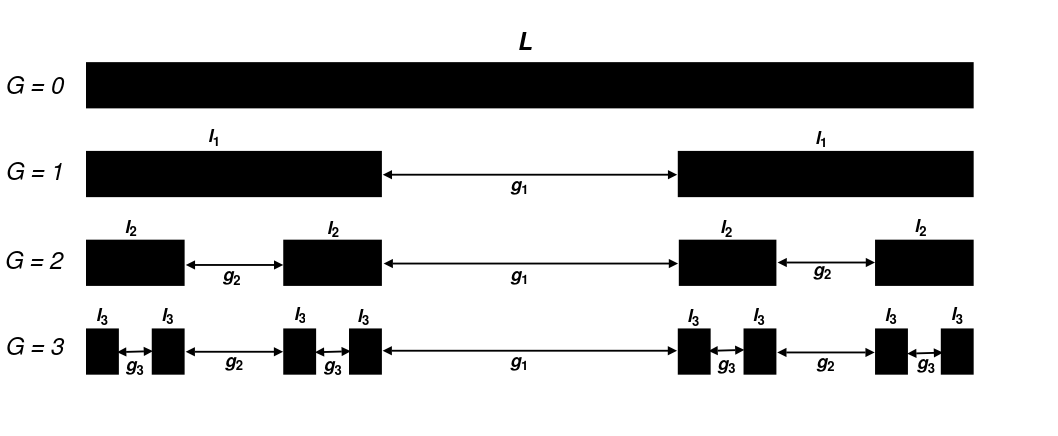}  
\caption{\textit{Symmetric Cantor family potential shows the length and gap between the segments. }}
\label{fractal_g}
\end{center}
\end{figure}
\\
From Fig. \ref{cons_sfp} and \ref{fractal_g}, we observe $s_{1}= l_{G}+g_{G}$, $s_{2}= s_{1}+ l_{G}+g_{G-1}$, $s_{3}= s_{1}+s_{2}+ l_{G}+g_{G-2}$, $s_{4}= s_{1}+ s_{2}+ s_{3}+ l_{G}+g_{G-3}$ and so on. Thus we arrive at,
\begin{equation}
s_{j}= \left ( \sum_{p=1}^{j-1} s_{p} \right ) + l_{G}+g_{G-j+1}.
\end{equation}
Therefore, $\eta_{1} (j)$ is given by,
\begin{equation}
\eta_{1} (j)=  -(l_{G}+g_{G-j+1}),
\label{eta1_lg}
\end{equation}
which shows that $\eta_{1} (j)$ is always negative. Combining Eq. \ref{eta2_jr} and \ref{eta1_lg} we get,
\begin{equation}
\eta_{2} (j,r)=  g_{G-r+1} - g_{G-j+1}.
\label{eta2_lg}
\end{equation}
We see from Fig. \ref{fractal_g} that for $i>j$, $g_{i}< g_{j}$. Also for $r<j$, $G-r+1 > G-j+1$ which implies $g_{G-r+1} < g_{G-j+1}$. Therefore $\eta_{2}(j,r)<0$ for $r<j$. Eq. \ref{eta1_lg} and \ref{eta2_lg} gives the general expression for $\eta_{1}$ and $\eta_{2}$ respectively. Now we provide these expressions for GC and GSVC cases. 


\subsection{Case 1: General SVC potential}
To calculate $\eta_{1,2}$ for GSVC, we re-write Eq. \ref{l_G_qp} as 
\begin{equation}
l_{j-1} = \frac{L}{2^{j-1}} q \left ( \frac{1}{\rho}; \frac{1}{\rho} \right)_{j-1}.
\end{equation} 
As we know, for GSVC a fraction $\frac{1}{\rho^{j}}$ is removed from segment length $l_{j-1}$ to generate the system for stage $G=j$,  therefore, $g_{j}= \frac{l_{j-1}}{\rho^{j}}$ and hence,
\begin{equation}
g_{j} = \frac{L}{\rho^{j} 2^{j-1}} q \left ( \frac{1}{\rho}; \frac{1}{\rho} \right)_{j-1}.
\label{gj_qp}
\end{equation} 
Now we simplify for $\eta_{1}(j)$ by using Eq. \ref{eta1_lg}, Eq. \ref{l_G_qp} and Eq. \ref{gj_qp} to obtain,
\begin{equation}
    \eta_{1}(j)=-\Biggl\{\frac{L}{2^{G}} q \left ( \frac{1}{\rho}; \frac{1}{\rho} \right)_{G} + \frac{L}{2^{G-j}} q \left ( \frac{1}{\rho}; \frac{1}{\rho} \right)_{G-j}\frac{1}{\rho^{G-j+1}}\Biggr\}.
    \label{eta_svc_rho_G}
\end{equation}
Now we calculate $\eta_{2}(j,r)$ by using Eq. \ref{eta2_jr} and \ref{eta_svc_rho_G} to obtain,
\begin{equation}
    \eta_{2}(j,r)= \frac{2L}{(2\rho)^{G+1}}\Biggl\{(2\rho)^{r}  q \left ( \frac{1}{\rho}; \frac{1}{\rho} \right)_{G-r}- (2\rho)^{j}  q \left ( \frac{1}{\rho}; \frac{1}{\rho} \right)_{G-j}\Biggr\}
    \label{eta2_svc_rho_G_1}
\end{equation}

Now  we can substitute Eq. \ref{eta_svc_rho_G} and \ref{eta2_svc_rho_G_1} in Eq.  \ref{zeta_j1} to obtain the general expression for `$\zeta_{j}$' for GSVC potential.

\subsection{Case 2: General Cantor potential}
We re-write Eq.  \ref{l_G_C} as,
\begin{equation}
l_{j-1}= \left(\frac{\rho - 1}{2\rho}\right)^{j-1}L.
\end{equation}
As we know, in case of GC potential, a fraction $\frac{1}{\rho}$ is taken from stage $G=j-1$ to create the fractal system for $G=j$ stage, therefore $g_{j}= \frac{l_{j-1}}{\rho}$ and thus,
\begin{equation}
g_{j}= \frac{1}{\rho} \left(\frac{\rho - 1}{2\rho}\right)^{j-1}L.
\label{gj_c}
\end{equation}
Now, using Eq. \ref{eta1_lg} and \ref{l_G_C} in Eq. \ref{gj_c}, we simplify for $\eta_{1}(j)$ to get,
\begin{equation}
\eta_{1}(j)= - \Biggl\{ L.\left(\frac{\rho-1}{2\rho}\right)^{G}+\frac{L}{\rho}.\left(\frac{\rho-1}{2\rho}\right)^{G-j} \Biggr\}.
\label{eta1_rhoG}
\end{equation}
Now using  Eq. \ref{eta2_jr}, $\eta_{2}(j,r)$ can be simplified as,
\begin{equation}
     \eta_{2}(j,r)=\frac{L}{\rho}\left(\frac{\rho-1}{2\rho}\right)^{G-r-j}\Biggl\{ \left(\frac{\rho-1}{2\rho}\right)^{j}-\left(\frac{\rho-1}{2\rho}\right)^{r} \Biggr\}. 
     \label{eta2_rhoG_1}
\end{equation}
Substitution of Eq. \ref{eta1_rhoG} and \ref{eta2_rhoG_1} in Eq. \ref{zeta_j1} gives the general expression of `$\zeta_{j}$' for GC potential. 
\section{Transmission features}
\label{Tras_Featu}
\begin{figure}[H]
\begin{center}
	\centering
\includegraphics[scale=0.90]{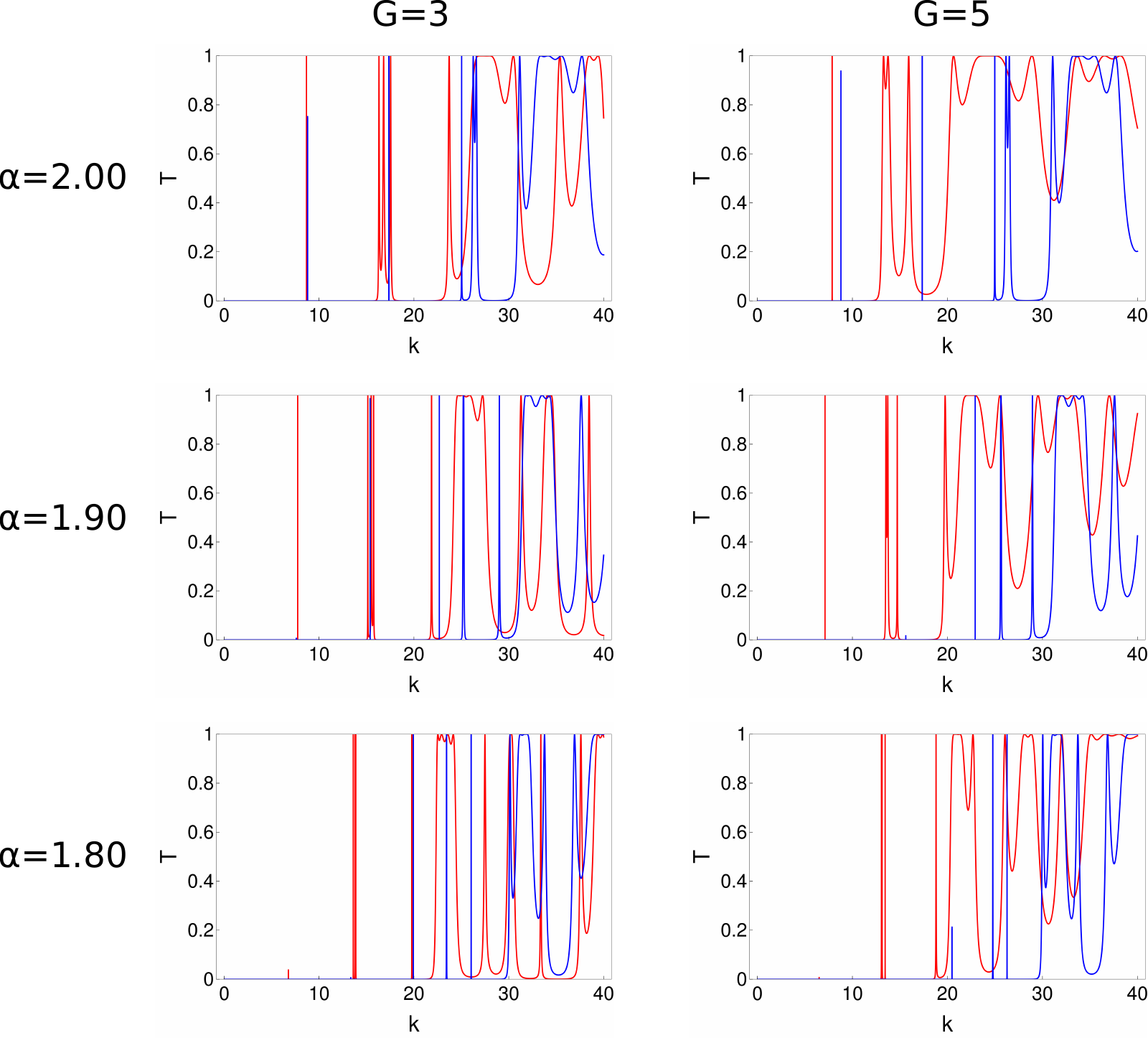}
\caption{\textit{Plots showing the comparison of transmission amplitudes for GC (Red-curve) and GSVC (Blue-curve) potential for two different stages of G (= $3$ and $5$) in SFQM ($\alpha$ = $2.00$, $1.90$ and $1.80 $). Here potential parameters are $L=1$, $V=400$ and $\rho=3.5$.  From figures, it is observed that GSVC potential has sharper peaks as compared to general Cantor potential. }}
\label{comparison_gc_gsvc_01}
\end{center}    
\end{figure}

In the previous section, the analytical expressions of the tunneling amplitudes from two types of Cantor potentials in SFQM have been derived. In this section, we study the various features of transmission through these systems in SFQM. As the Cantor potentials have been studied in detail in standard QM ($\alpha = 2$) \cite {cantor_f1, cantor_f2,cantor_f3,cantor_f4,cantor_f5,cantor_f6,cantor_f7,cantor_f7_1,cantor_f8,cantor_f9,Hasan_SPP}, therefore we largely focus here to study the tunneling behavior in the domain of SFQM (i.e., the case of $\alpha < 2$) as well as the comparison with the case of standard QM (i.e., the case of $\alpha = 2$). Fig. \ref{comparison_gc_gsvc_01} shows the comparison of the profiles of the transmission amplitudes for GC and GSVC potential in standard QM and in SFQM for different stages G. In all plots of Fig. \ref{comparison_gc_gsvc_01}, it is noted that the transmission resonances are much sharper in GSVC potential as compared to GC potential. As these two types of potentials are different, they show different transmission profiles which are not relatable (at the present level of investigations) though both are special cases of repeating systems. Therefore, we present these two cases separately in subsequent sections. Subsection \ref{general_svc_features} discusses the case of GSVC while subsection \ref{general_cantor_features} details the case for GC potential. 

\subsection{Transmission features of general SVC potential in SFQM}
\label{general_svc_features}
\begin{figure}[H]
    \centering
    \includegraphics[scale=0.65]{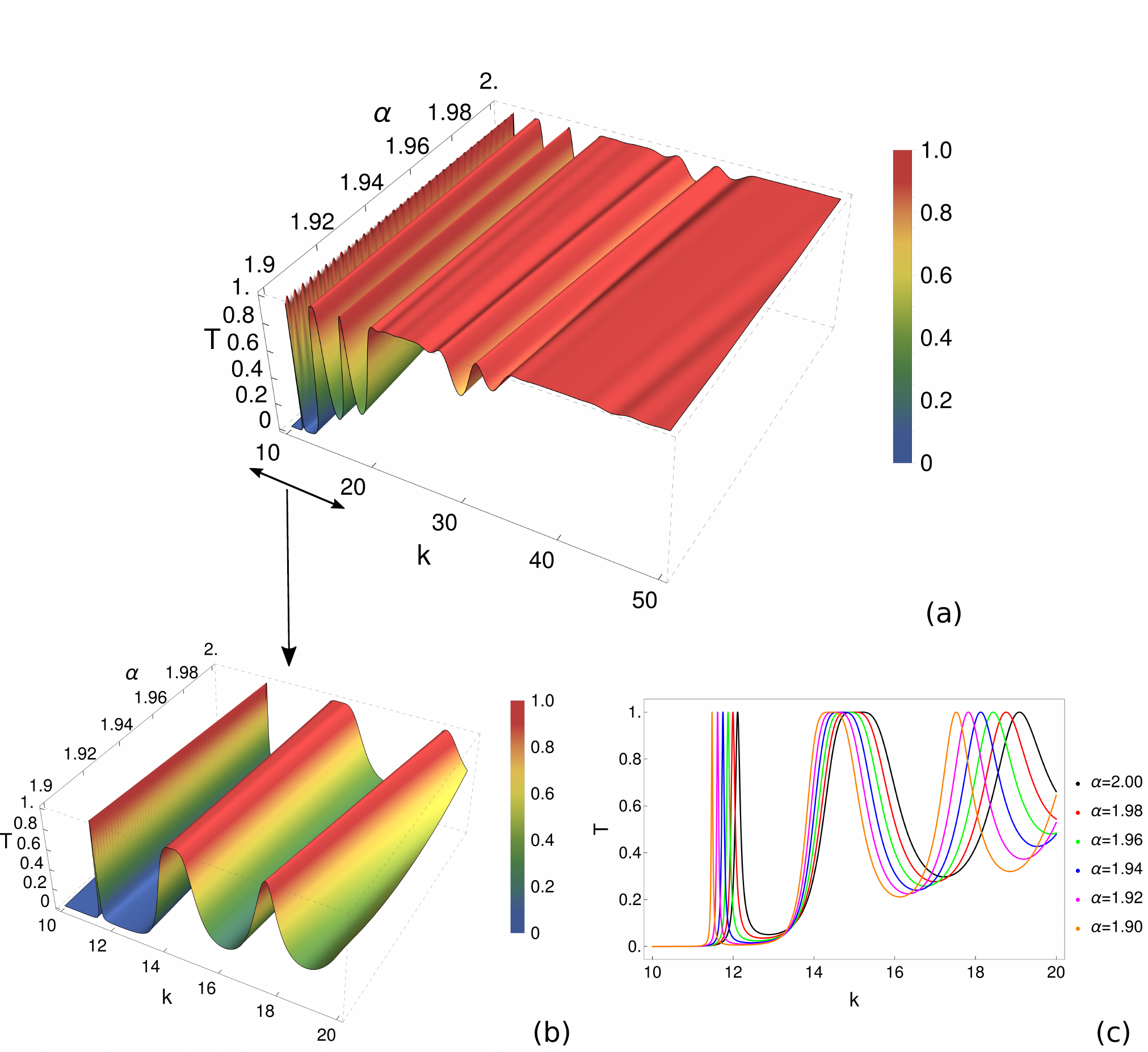}
    \caption{\textit{The transmission amplitude for GSVC potential in SFQM with $\alpha$ and $k$. The potential parameters are $V=100$, $\rho=3$ and $G=3$. The transmission peaks occurs at lower $k$ values with decreasing $\alpha$. It is also evident from Fig. (c) that the sharpness of the transmission peaks are increasing as $\alpha$ is lowered. }}
    \label{gsvc_comb}
\end{figure}
\begin{figure}[H]
	\begin{center}
		\includegraphics[scale=0.35]{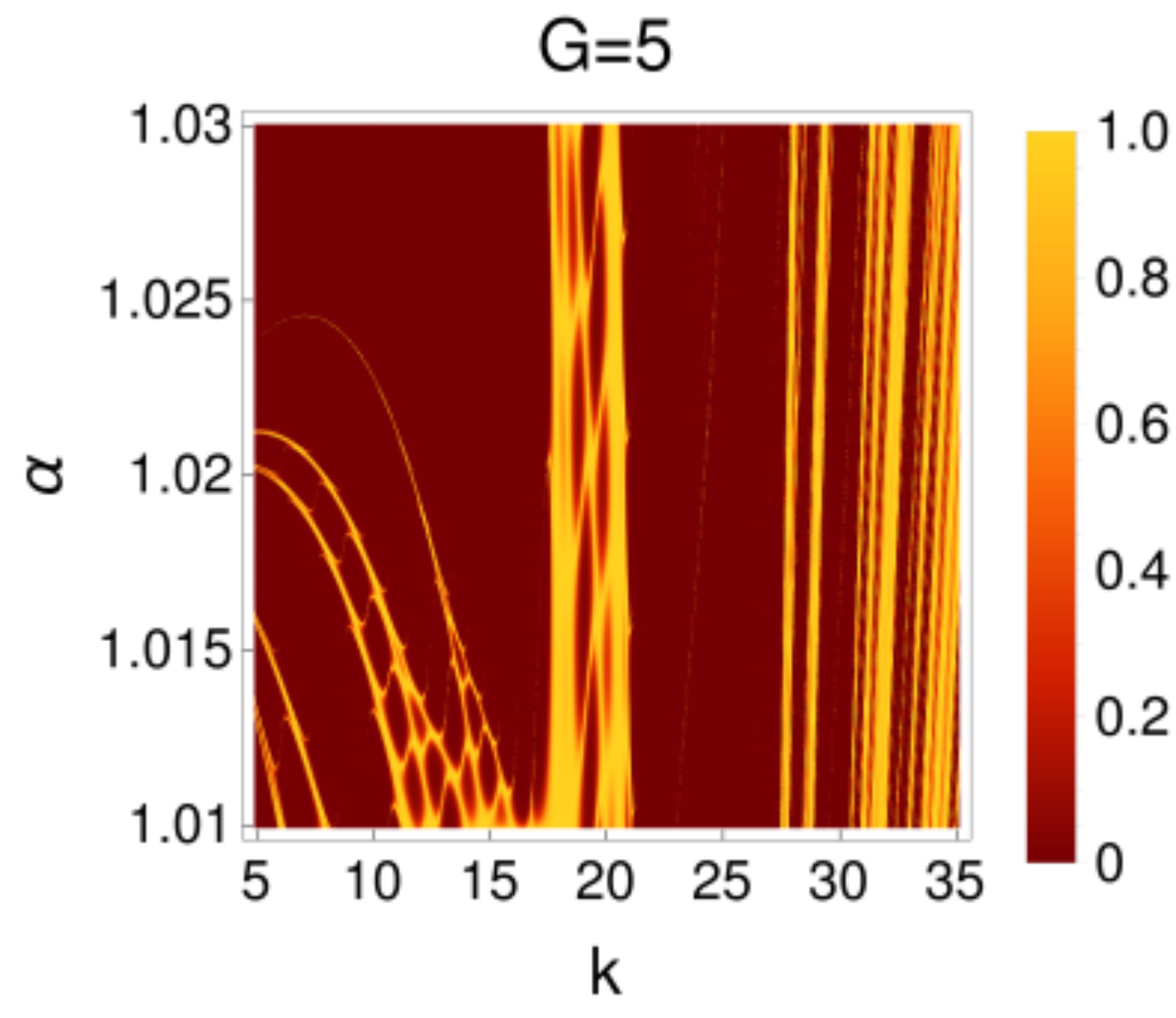} (a)
		\includegraphics[scale=0.35]{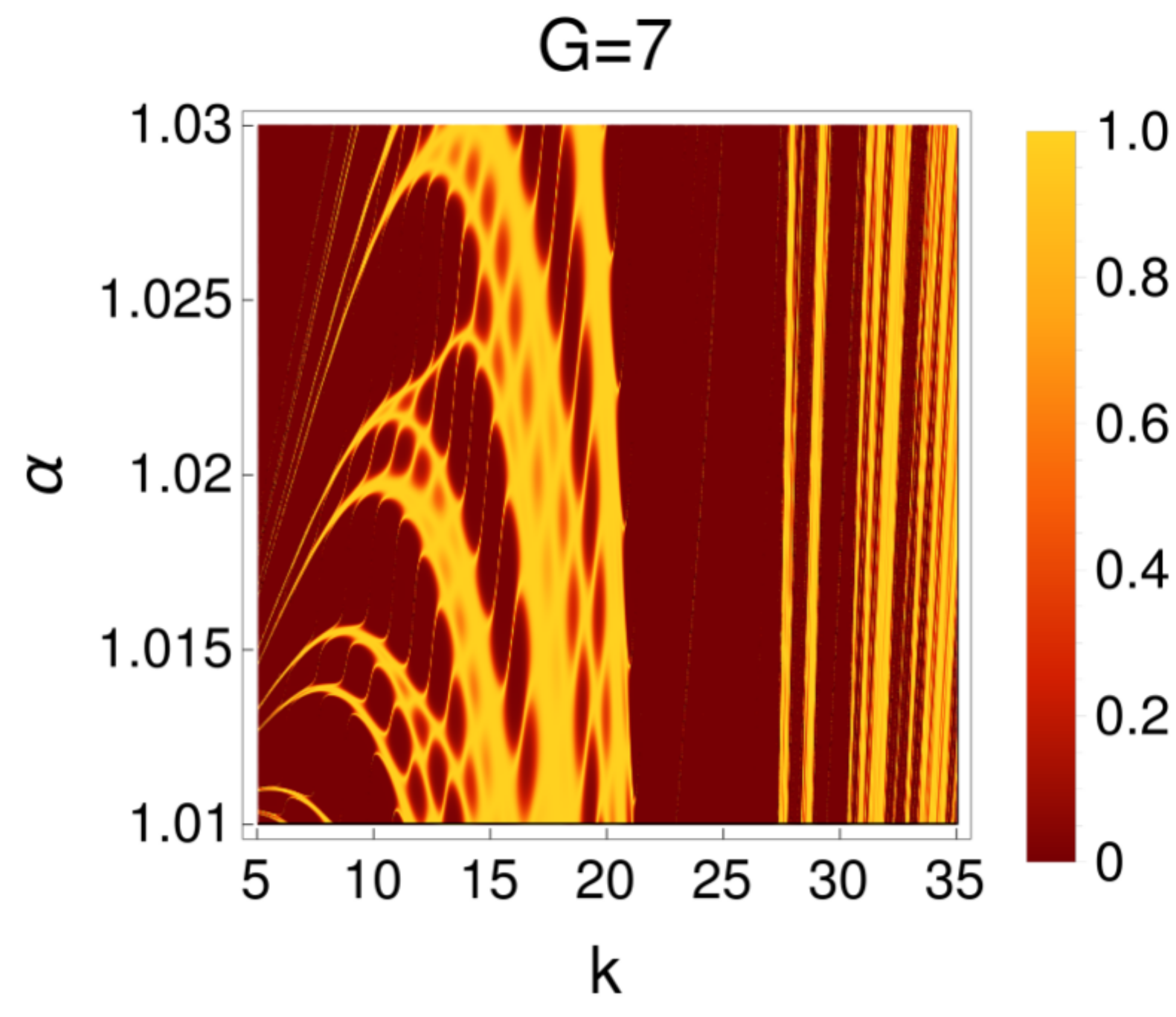} (b)
		\caption{\textit{Density plot showing the variation of transmission amplitude $T$ in $\alpha -k$ plane for $\alpha$ close to 1 for GSVC potential of different stages $G=5$ and $7$. The potential parameters are $V=300$, $L=1$ and $\rho=3$. Extreme sharp transmission resonances are seen. For both the stages of the potential,  extreme behavior of variations in $T$ is observed for wave energy $E< V$. However, for $E > V$, the transmission amplitudes are observed to saturate with increasing $G$. Further, deep minima in $T$ occur in the transmission profile for several finite ranges of $k$ which indicates the presence of allowed and forbidden band-like structures from this potential system in SFQM.}}
		\label{gsvc_04}
	\end{center}    
\end{figure}
\begin{figure}[H]
\begin{center}
    \centering
    \includegraphics[scale=0.70]{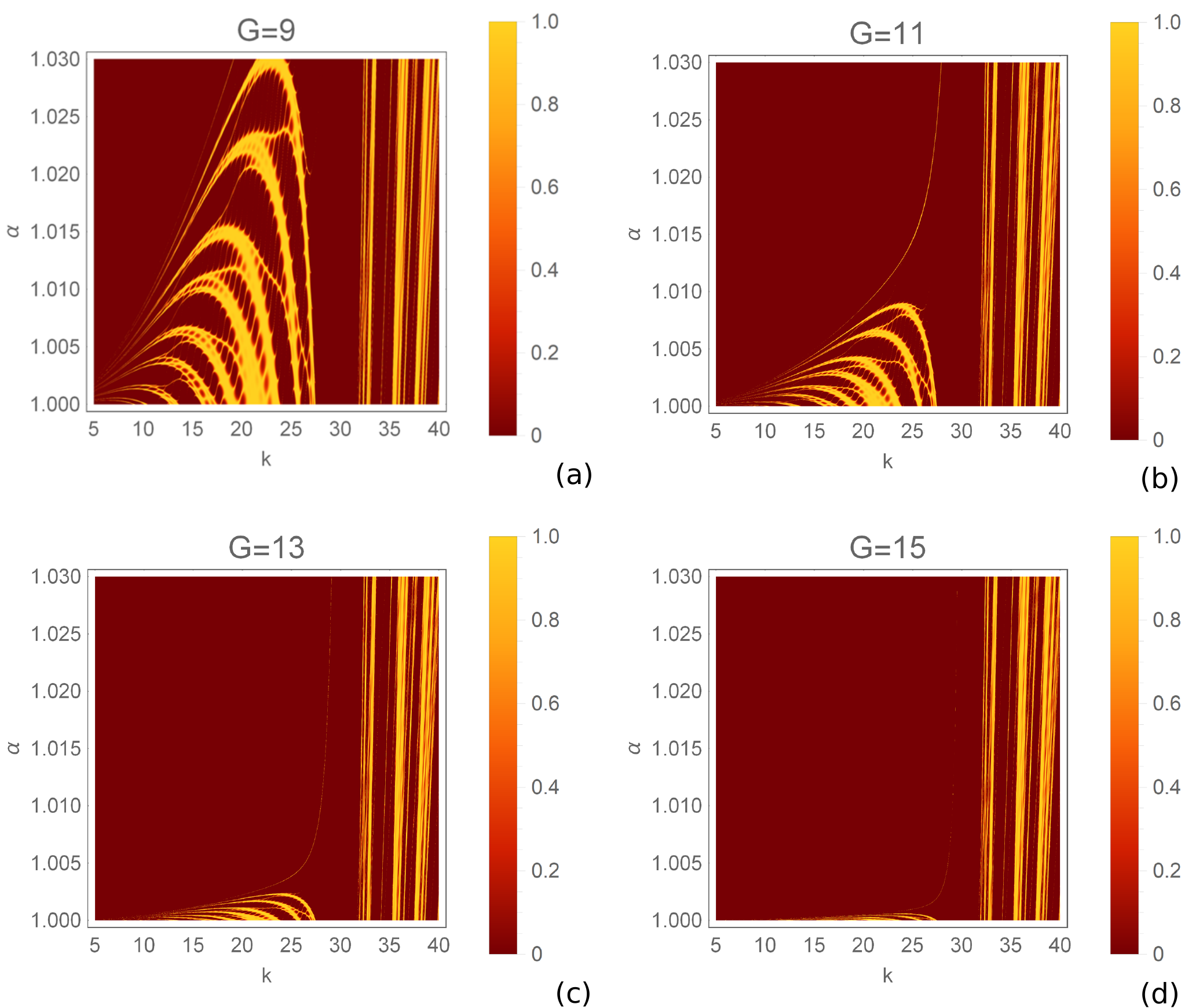}
    \caption{\textit{Density plot showing the variation of transmission amplitude $T$ in $\alpha -k$ plane for $\alpha$ close to 1 for GSVC potential of different stage $G$. Here  $V=450$ and other parameters are the same as Fig. \ref{gsvc_04}. Again, the presence of extremely sharp transmission resonances is noticed with extreme variations in $T$ for wave energy $E<V$. It is also seen in the figure that the transmission amplitude saturates with increasing $G$ for $E>V$. Again, the density plot shows the occurrence of band-like features.}}
\label{gsvc_05}
\end{center}    
\end{figure}

This section exclusively discusses the nature of the transmission profile from GSVC system in SFQM. As the expression for the transmission amplitude is transcendental in nature (Eq. \ref{tg_n1n2ng}), presently we rely on the numerical investigation towards investigating the general features of tunneling amplitude. The transmission amplitude is plotted for stage $G=3$ in Fig. \ref{gsvc_comb}. To understand the behavior of transmission resonances with $\alpha$, we 3D plot $T(\alpha,k)$ with $\alpha$ and $k$ as shown in Fig. \ref{gsvc_comb}-a. Here the potential parameters are $V=100$, $\rho=3$, and $G=3$. A closer look at this figure is shown in Fig. \ref{gsvc_comb}-b for a smaller range of $k$. From both these figures, it is seen that the locus of transmission resonances has a positive slope with increasing $\alpha$. This indicates that the transmission peaks are red-shifted with decreasing values of $\alpha$. This appears to be a general trend for the case when $\alpha$ is not far away from $2$. However, much more complex behavior of the locus of transmission resonances is seen when $\alpha$ is closer to $1$ and is presented later in the paper. Fig. \ref{gsvc_comb}-c shows the $2$D plot depicting the variation of $T$ for different $\alpha$ with the same range of $k$ as shown in Fig. \ref{gsvc_comb}-b. This figure shows that the transmission resonances become sharper at lower values of $\alpha$. This appears to be a general feature and will be more evident in the later part of the discussion and associated graphical representations.      
\begin{figure}[H]
\begin{center}
\includegraphics[scale=0.32]{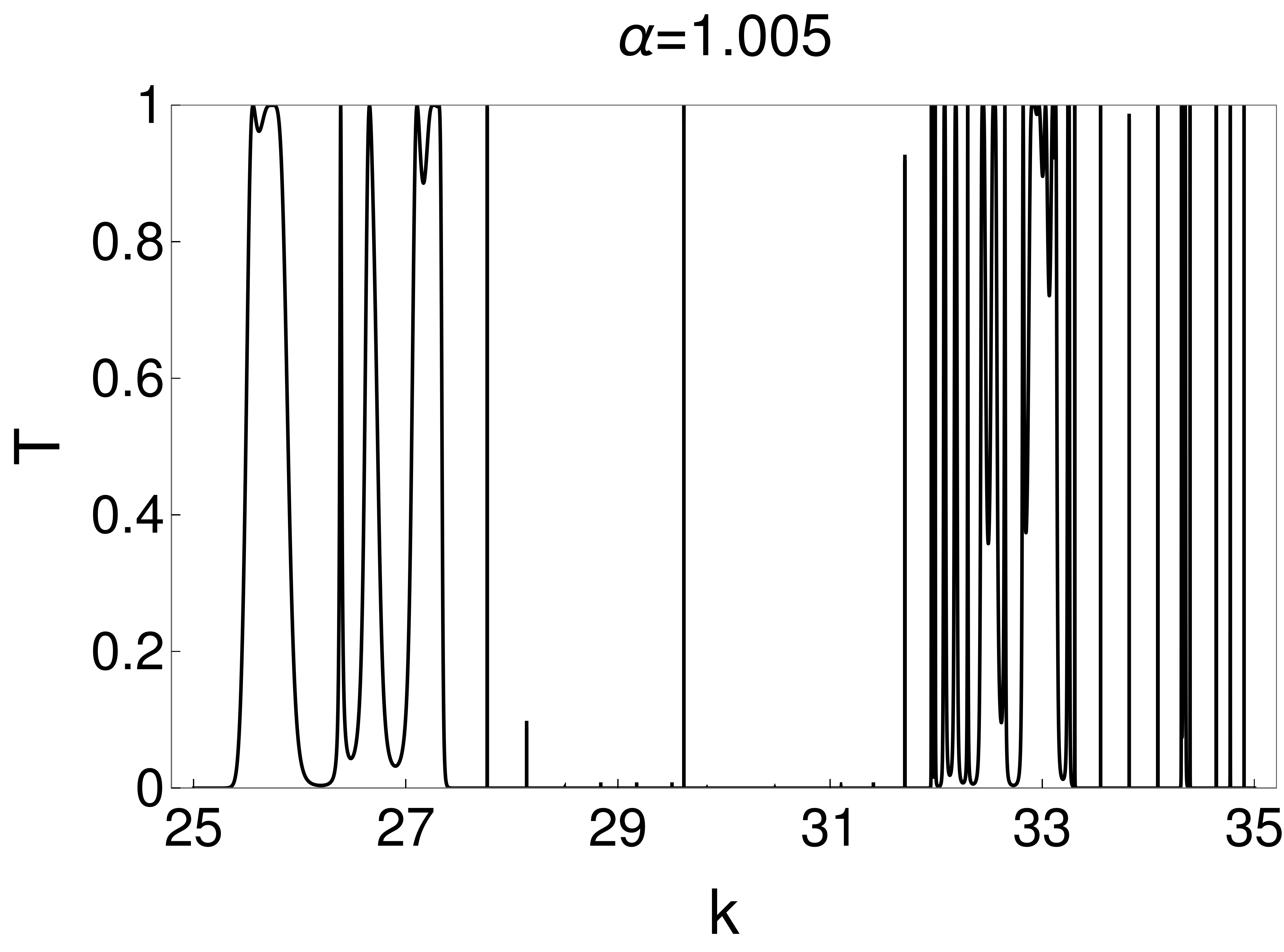} (a) \ \includegraphics[scale=0.32]{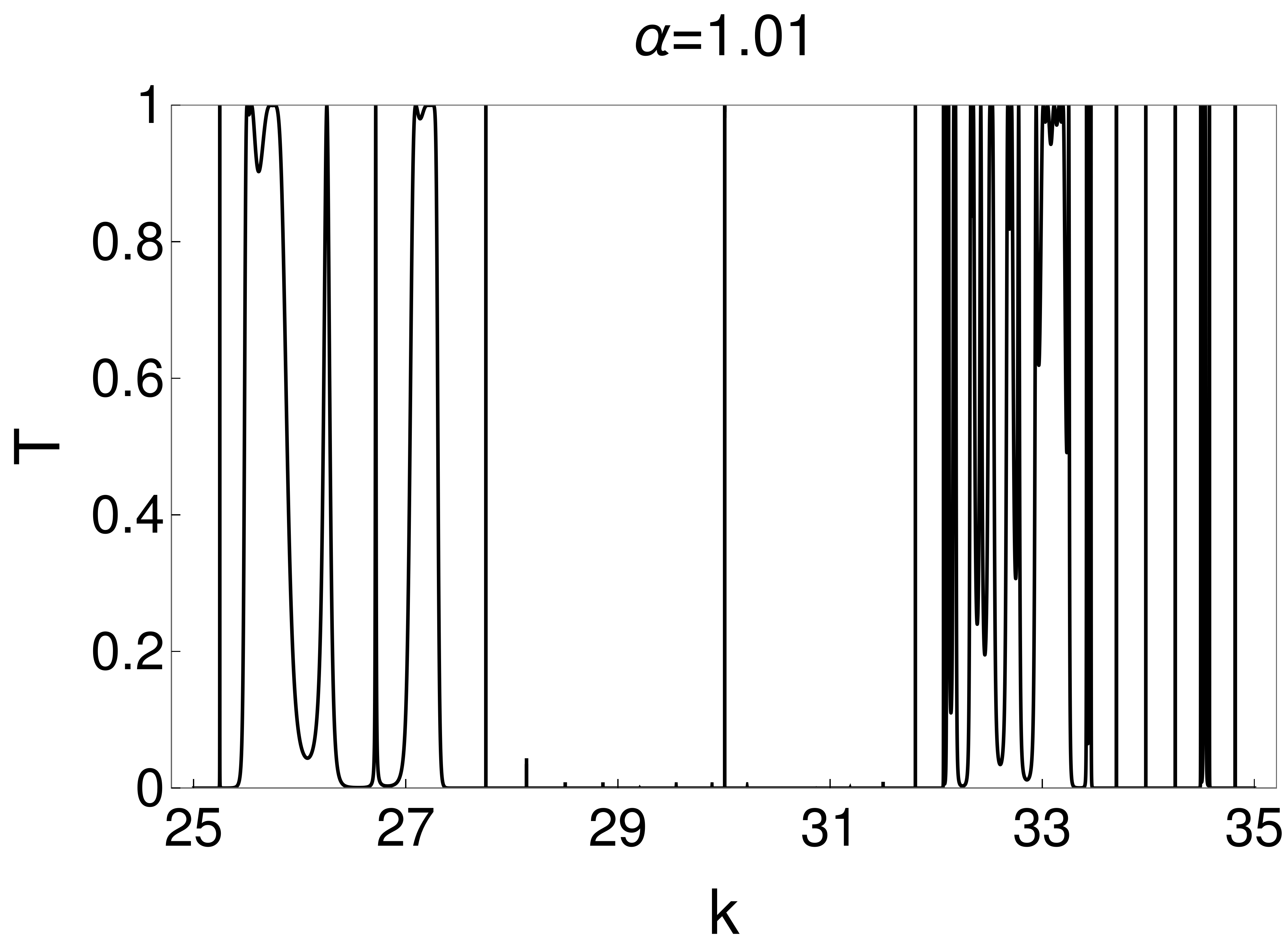}  (b) \\
\includegraphics[scale=0.32]{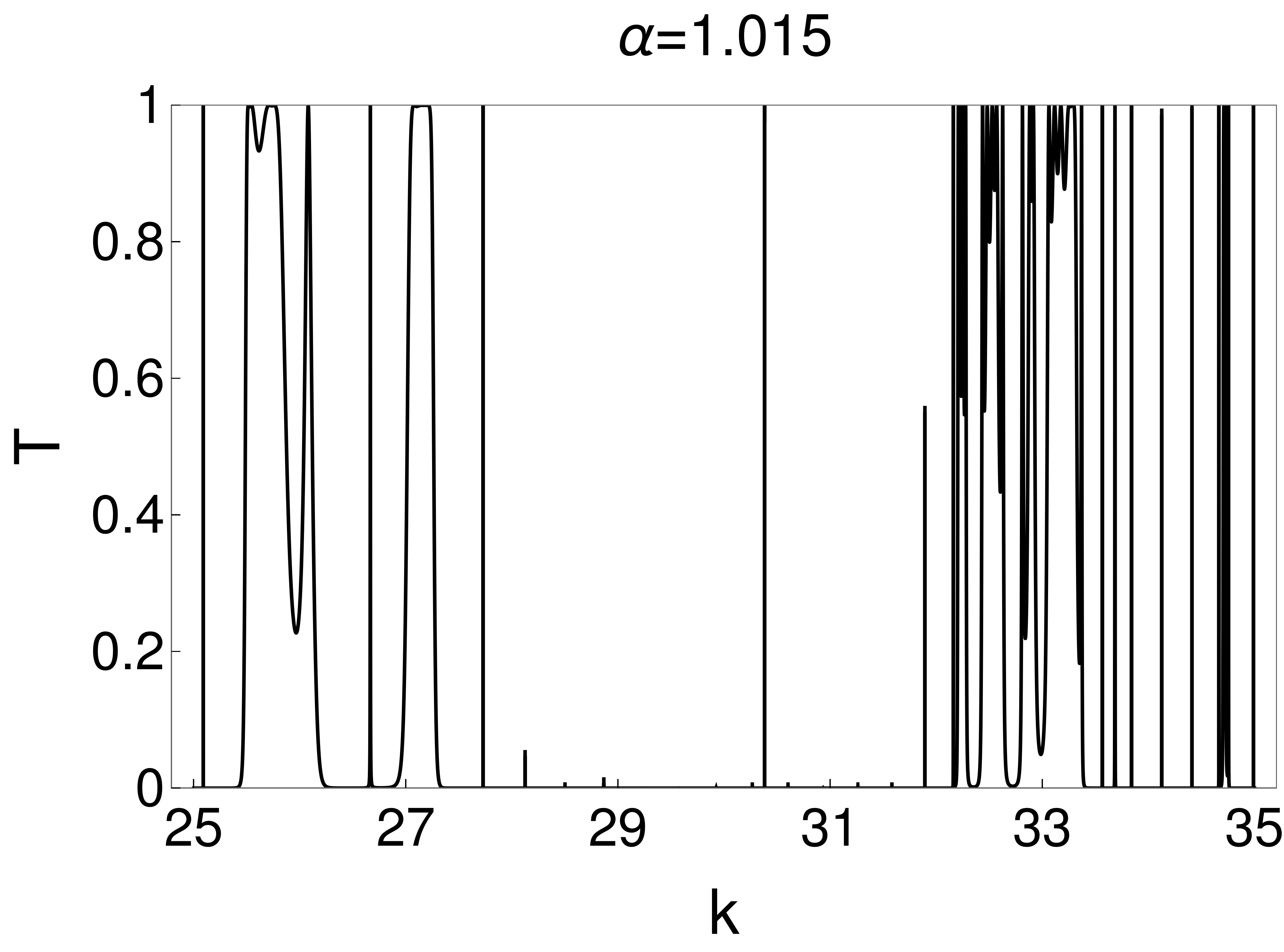} (c) \ \includegraphics[scale=0.32]{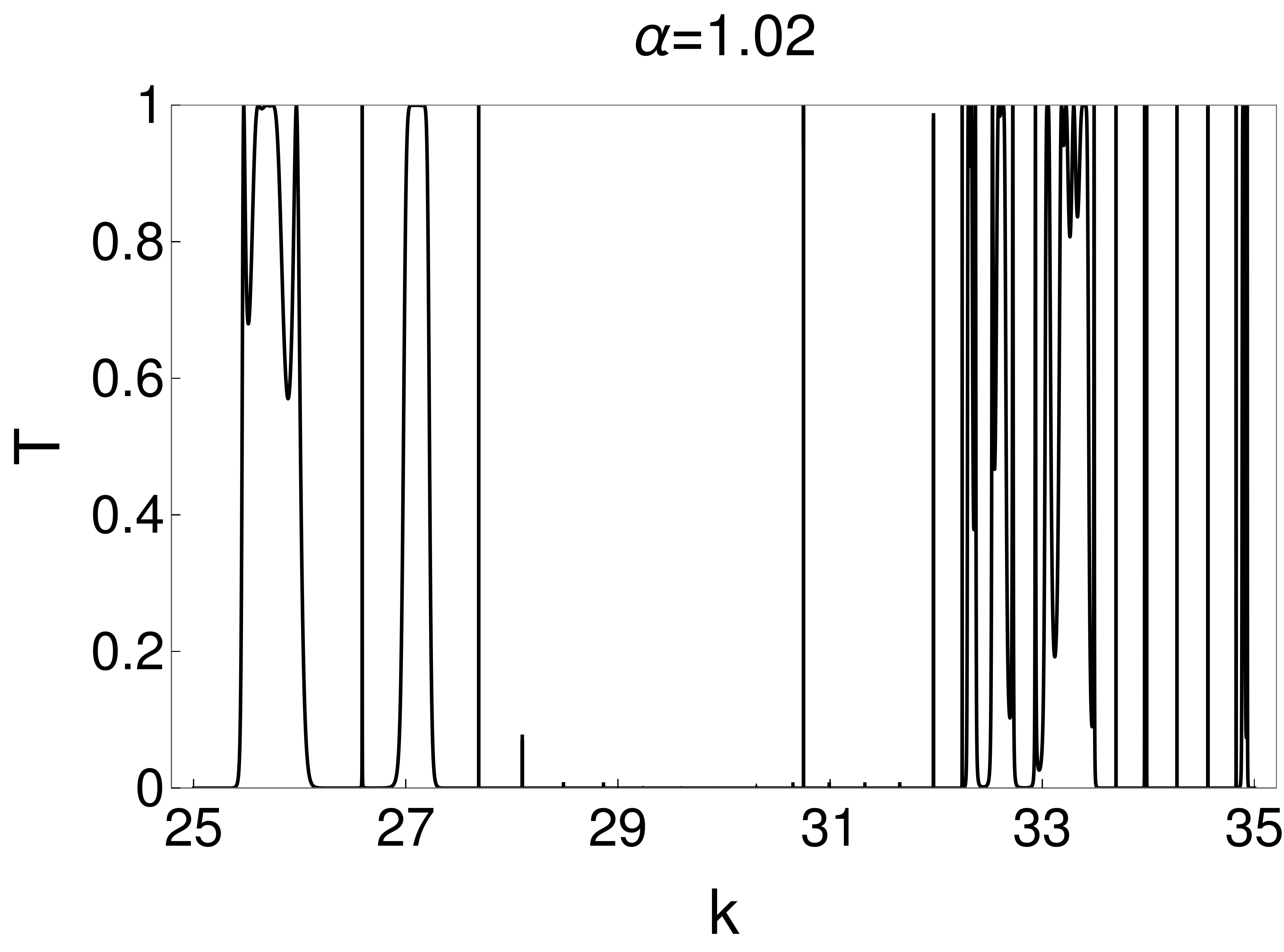}  (d) \\
\includegraphics[scale=0.32]{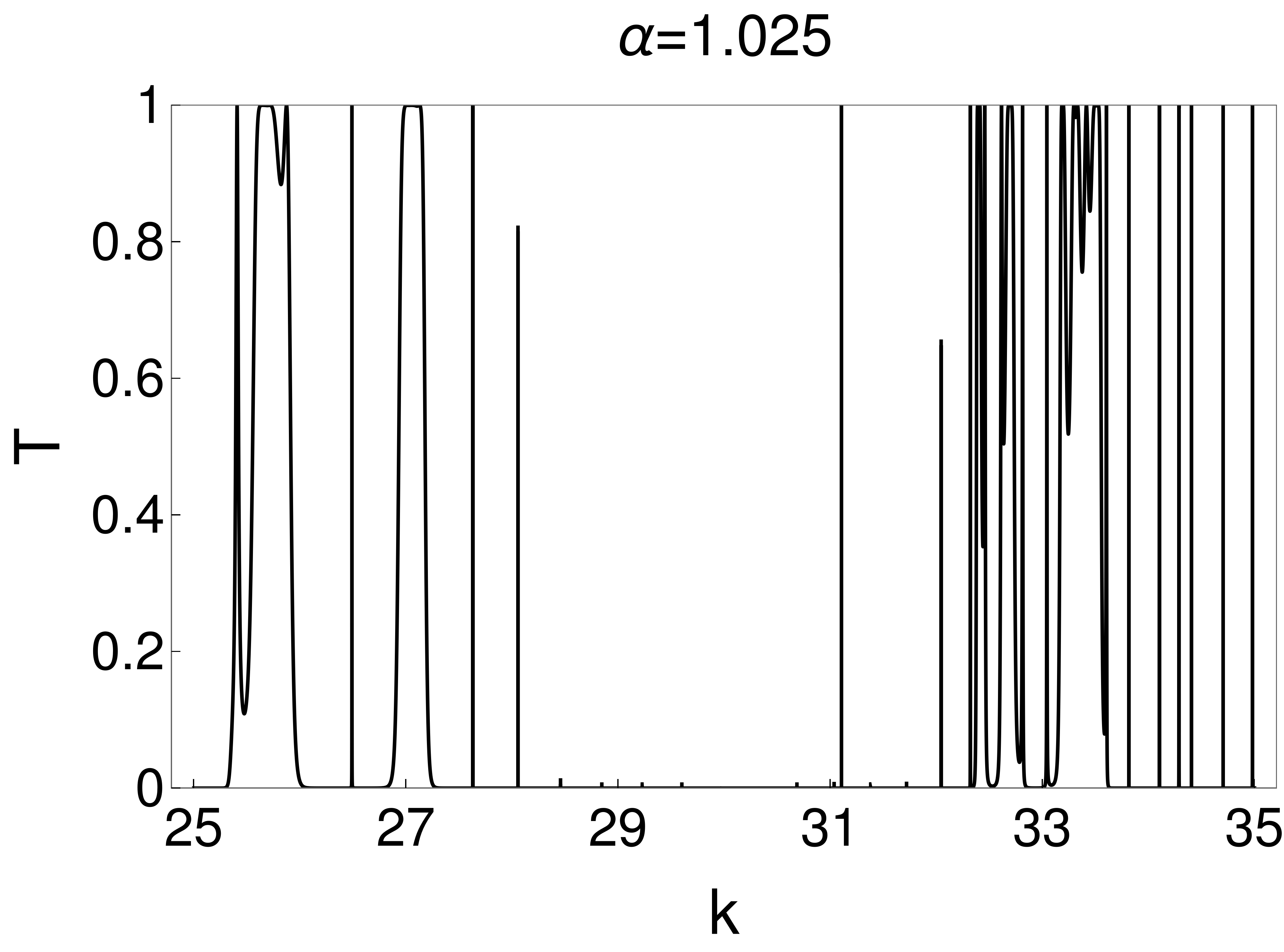} (e) \ \includegraphics[scale=0.32]{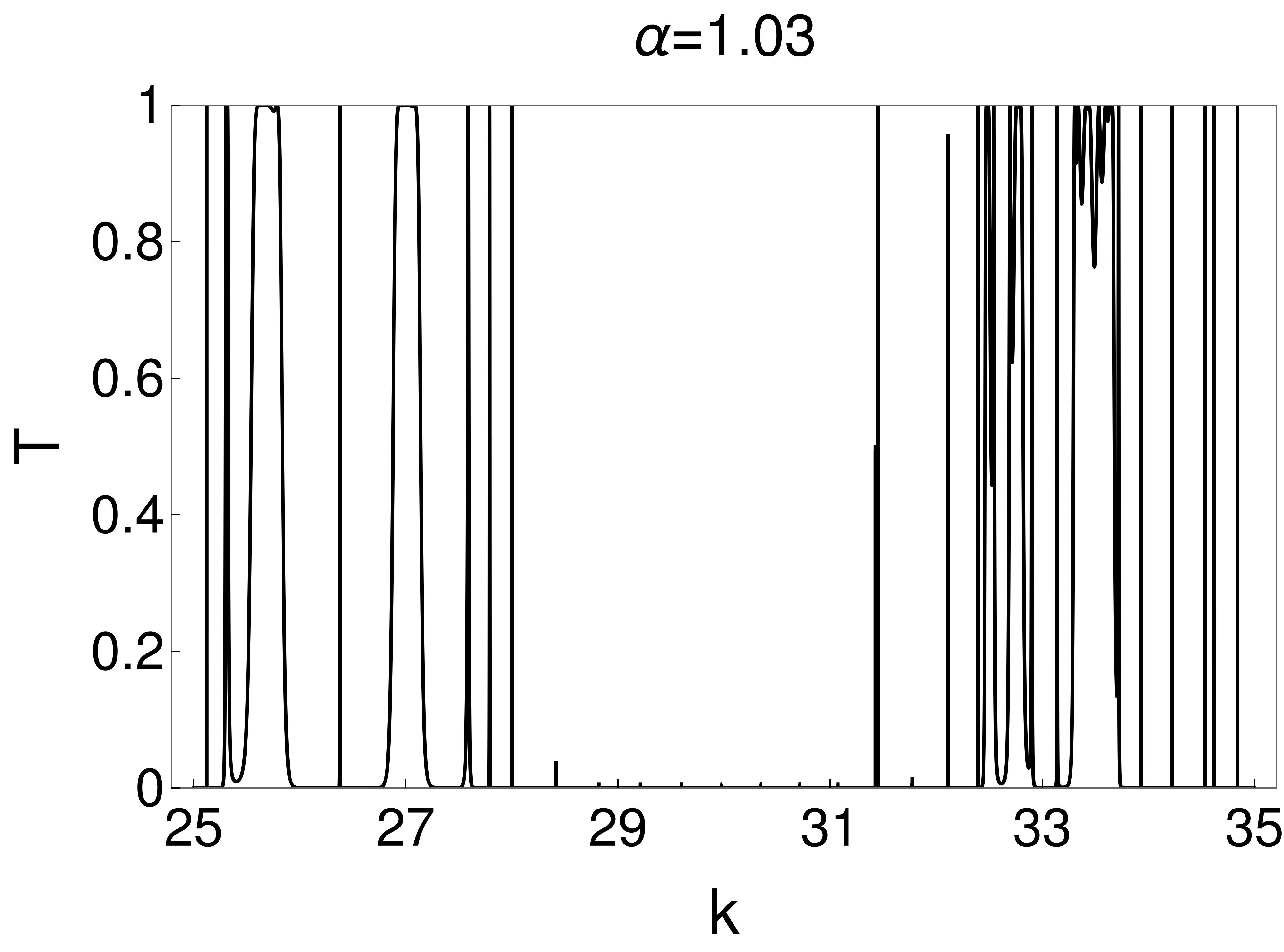}  (f) \\
\caption{\textit{Plots showing several sharp transmission resonances near unity (i.e $\alpha =1$) for GSVC potential of stage $G=5$ in SFQM. The potential parameters are $V=450$, $L=1$ and $\rho=3$.}}
\label{gsvc_02}
\end{center}    
\end{figure}
\paragraph{}
An interesting parameter region for the study of tunneling amplitude in SFQM is the case when $\alpha$ is close to $1$. In this regime, extreme behavior in the transmission amplitudes is observed which is demonstrated graphically in Fig. \ref{gsvc_02}. The figure shows the transmission amplitude for stages $G=5$, $V=450$, $L=1$, and different values of $\alpha$ near unity. In all these figures, the emergence of several extremely sharp transmission resonances is observed for both evanescent and non-evanescent waves. The transmission resonances are separated by deep valleys in the $T(k)$ profile such that $T(k)$ vanishes over a range of $k$ (it may be noted that for any Hermitian potential, as in the present case transmission amplitudes are never ideally zero \cite{Mostafazadeh2018}). Many transmission resonances in Fig. \ref{gsvc_02} are extremely sharp and appear as the sudden jump from $T=0$ to $T=1$. Towards understanding these features over a continuous range of $\alpha$ near $1$, the transmission amplitudes are represented through density plots in $\alpha-k$ plane for different stages of the potential in Fig. \ref{gsvc_04} and Fig. \ref{gsvc_05}. For both these figures $L=1$, $\rho=3$ while $V=300$ and $450$ for Fig. \ref{gsvc_04} and Fig. \ref{gsvc_05} respectively.

\begin{figure}[H]
    \centering
    \includegraphics[scale=0.65]{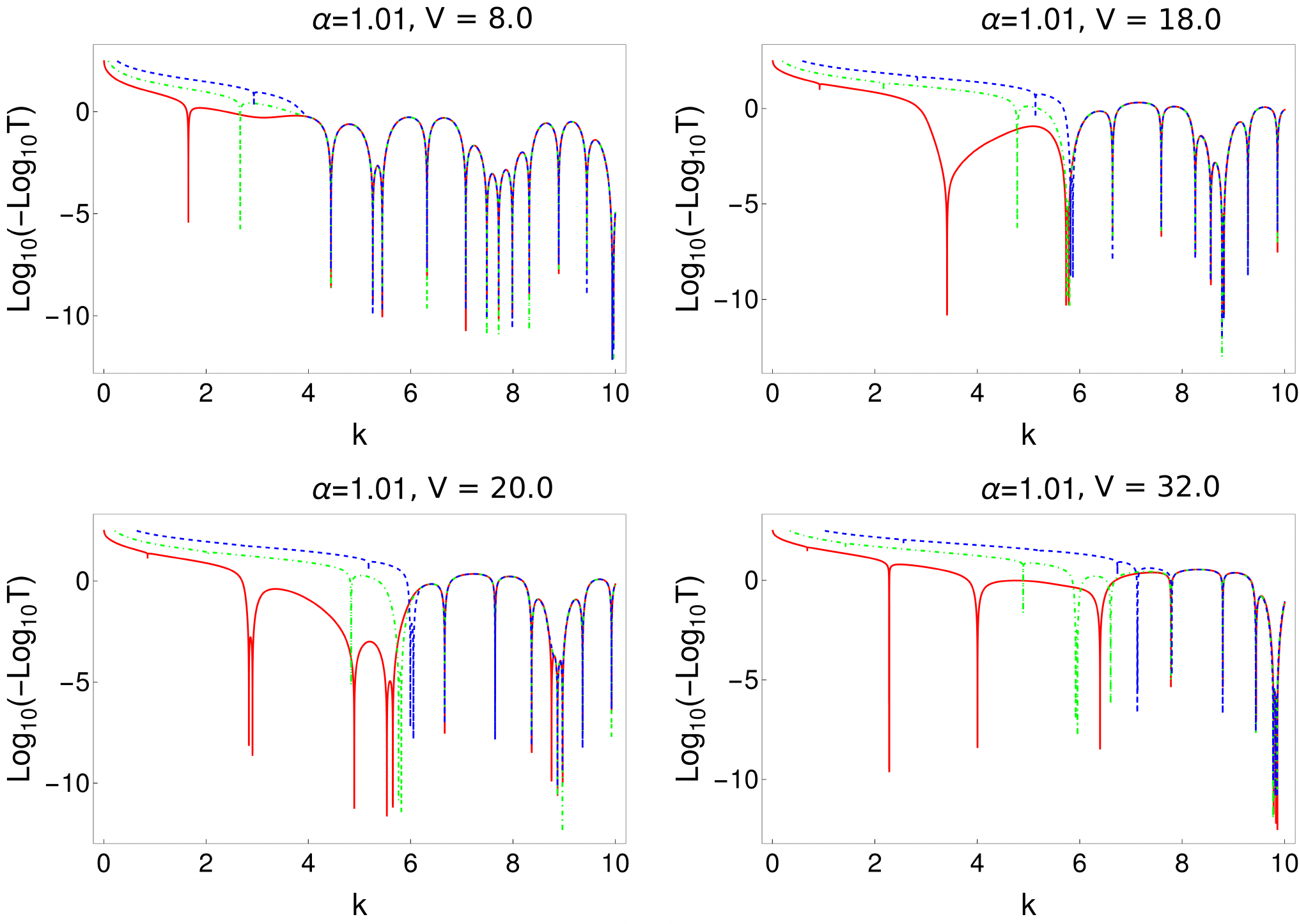}
    \caption{\textit{Plot of $\log_{10}{(-\log_{10} {T})}$ for the case of GSVC  for $G=5$ (red curve), $G=8$ (dashed green curve), and $G=11$ (dashed blue curve). The potential parameters are $L=1$, $\rho=3$, and the height $V$ is indicated in each figure. It is seen from all these plots that for values of $\alpha$ close to $1$, $T(k)$ profile saturates with $G$ when wave energy $E (= k^{2}/2m)>V$. No saturation in the $T(k)$ profile is observed when $E<V$. }}
    \label{saturation_with_ev}
\end{figure}

The different stages $G$ are shown in the figures. Both these figures indicate the presence of extremely sharp transmission resonances as thin streaks of yellow lines.  In some cases, the lines are so thin that these are not captured graphically over the red regions. The behavior of these $T=1$ loci is challenging to understand analytically due to the transcendental nature of the expression of the tunneling amplitudes. Extreme variations for both $\alpha$ and $k$ are observed for wave energy $E<V$ while for $E>V$ the transmission profile appears to saturate with increasing $G$. An apparent conclusion that may be drawn from Fig. \ref{gsvc_04} and \ref{gsvc_05} is the presence of deep valleys in the transmission amplitudes for which $T$ is nearly vanishing in $\alpha-k$ plane for $\alpha$ in the vicinity of $1$. The presence of these valleys in the transmission amplitudes is noted as the precursor of the emergence of allowed and forbidden energy bands  \cite{griffiths1992scattering} from locally periodic potential. This shows that the band features emerge in the case of GSVC potential in SFQM. In the case of standard QM, the band doesn't appear from Cantor family potential to the best of our knowledge. However, the emergence of band-like features from GSVC potential (and will be shown later for GC potential) appears only when $\alpha$ is in the vicinity of $1$. 
\paragraph{}
A discussion is in order. The deep valleys in $T(k)$ profile for a periodic potential for a range $\Delta k$ means that the waves are reflected from the potential for $k \in \Delta k$.  From the emergence of deep valleys in $T$ for tunneling through locally periodic delta potential, it is argued that the band-like structures emerge even when number of periodic delta barriers are just five, $N=5$ \cite{ griffiths1992scattering, Griffiths_wave}. Based on the similar studies in SFQM, it is noted that the band emerges even when $N=4$ and are more prominently present for lower $\alpha$ values \cite{tare}. In the present case, the repetitions are based on $N_{i}=2$. As $N=2$ system doesn't show band structure (and deep valleys in $T$ for ranges of $k$) in standard QM, the type of Cantor family system studied here don't show allowed energy bands in standard QM. 
\begin{figure}[H]
	\begin{center}
		\centering
		\includegraphics[scale=0.65]{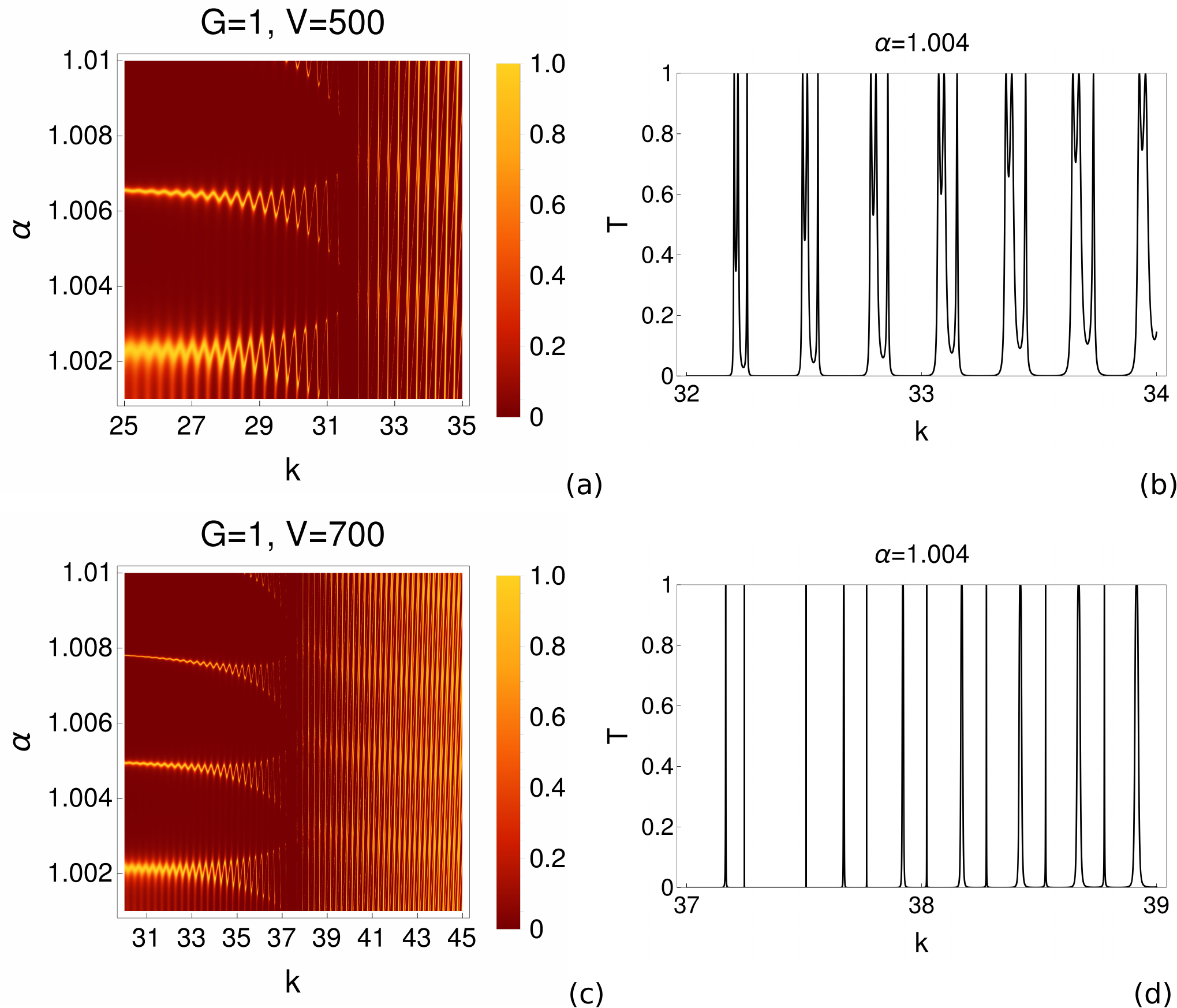}
		\caption{\textit{Plot of transmission amplitude for $G=1$ (double barrier system), above Fig. (a) shows density plot for potential height $V=500$ and Fig. (b) represent 2D plot for potential height $V=648$ with potential width $L=1$ for both cases. Both these plots show the occurrence of valleys for which $T$ is very close to zero. Fig. (c) shows the density plots for potential parameters $V=700$, $L=1$ in $\alpha-k$ plane for range of $\alpha$ from $1.001$ to $1.01$ and Fig. (d) corresponding 2D plot when $\alpha=1.004$ for same potential parametrs as Fig. (c) . The density plot clearly shows range of $k$ for which $T$ nearly vanishes. This again indicates the presence of allowed bands for the double barrier system. Interesting oscillations are seen in $T$ over $\alpha -k$ plane for the evanescent waves.}}
		\label{band_g1}
	\end{center}    
\end{figure}
Thus, a question may arise that if the present GSVC system, which is an arrangement of $N_{i}=2$ barriers, show band likes features for $\alpha \rightarrow 1$, would this also mean that a double barrier system will show  energy bands like features for  $\alpha \rightarrow 1$. Surprisingly, we find that this is indeed the case and are graphically shown in Fig. \ref{band_g1} for three different double barrier potential systems in SFQM. It is an extraordinary fact to recognize that there are allowed and forbidden bands for just double barrier systems in the domain of SFQM. If the $N=2$ barriers system could show band structures in SFQM, therefore the present GSVC systems which are  $N_{i}=2$ repeating systems could also show energy band structures. We will show in the later section that this is also true for GC potential in SFQM for $\alpha$ near to 1.      
\begin{figure}[H]
    \centering
    \includegraphics[scale=0.50]{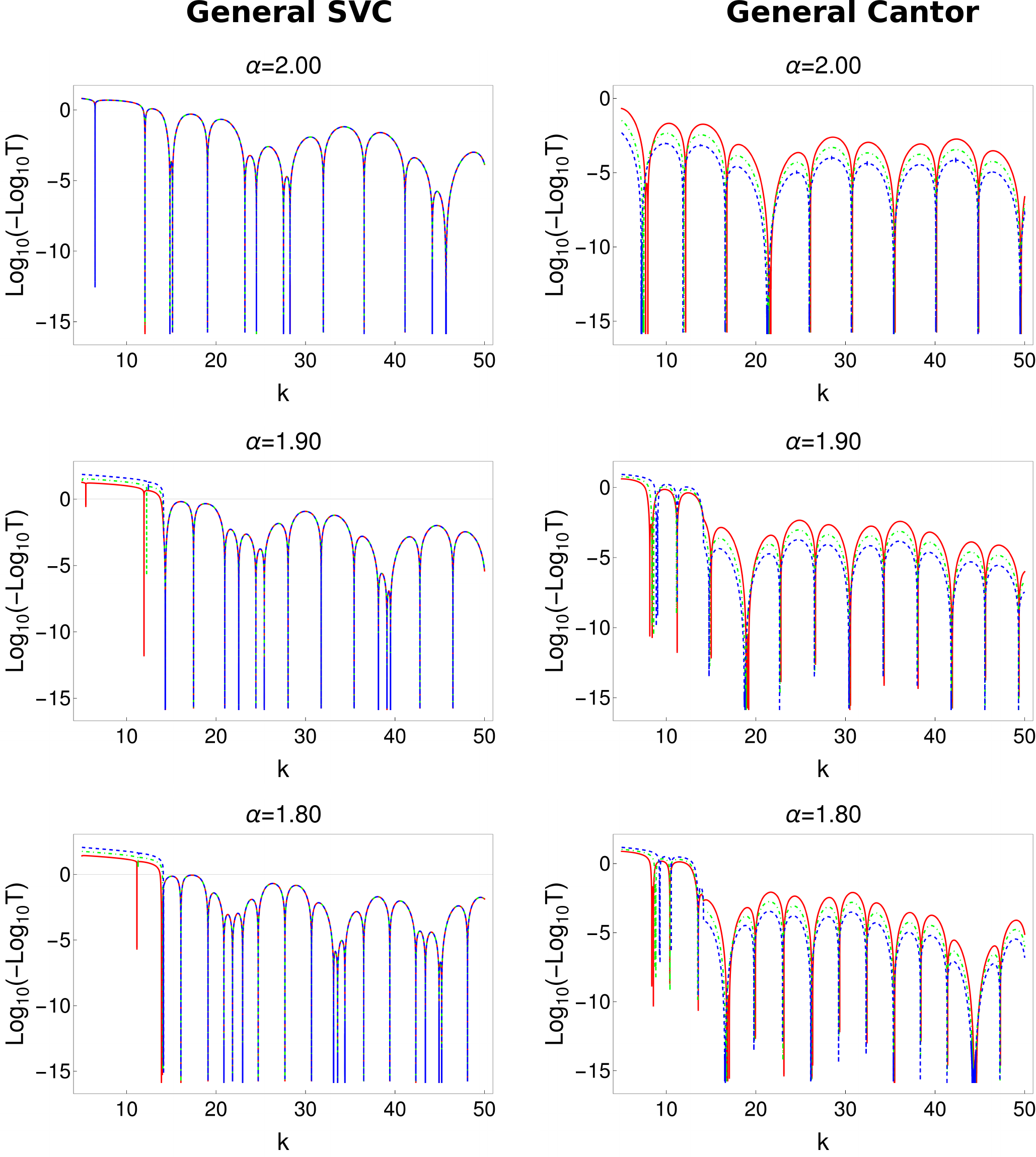}
    \caption{\textit{Plot of $\log_{10}{(-\log_{10} {T})}$ for the case of general SVC and general Cantor potential for $G=7$ (red curve), $G=9$ (dashed green curve), and $G=11$ (dashed blue curve). The potential parameters are $V=100$, $L=1$ and  $\rho=3$. As it is clearly visible from first column (for general SVC) and second column (for general Cantor), that the tunnelling saturates with increasing $G$ in standard (i.e., $\alpha=2$) as well as in SFQM for general SVC potential. However, this saturation behavior is not observed for general Cantor potential in SFQM.}}
    \label{gsvc_saturation_with_g}.
\end{figure}
\paragraph{}
Another observation from Fig. \ref{gsvc_05} is the saturation of the transmission profile with increasing $G$. From the definitions of GSVC system, a portion $\frac{1}{{\rho}^{G}}$ is taken out from the middle at each stage $G$. Thus progressively lesser fractions are taken from each stage with increasing $G$. This would imply that for larger $G$, the transmission profile should saturate with $G$ as only very thin portions are removed from the segments of the previous stages when $G$ is large. This is illustrated graphically in Fig. \ref{gsvc_saturation_with_g} in which a function of $T$ is plotted for GSVC and GC  potential for stages $G=7, 9$ and $11$ for different $\alpha$. For a better resolution in different $T(k,G)$, we have plotted $y=\log_{10}{(-\log_{10} {T})}$ in $y$-axis. As $0 < T \leq 1 $, therefore $\log_{10}{T} \leq 0$ and thus $-\log_{10}{T} \geq 0$. This implies that the function $y=\log_{10}{(-\log_{10}{T})}$ is well defined. Various plots with different $\alpha$ in Fig. \ref{gsvc_saturation_with_g} show that the saturation in $T(k)$ profile with $G$ is observed in GSVC potential but not in GC potential. However, for $\alpha \rightarrow 1$, this saturation is present only in the case for wave energy $E>V$ and not for $E<V$ as shown in Fig. \ref{saturation_with_ev} for different potentials and $\alpha=1.01$.
 
\subsection{Transmission features of general Cantor potential in SFQM}
\label{general_cantor_features}
\begin{figure}[H]
    \centering
    \includegraphics[scale=0.65]{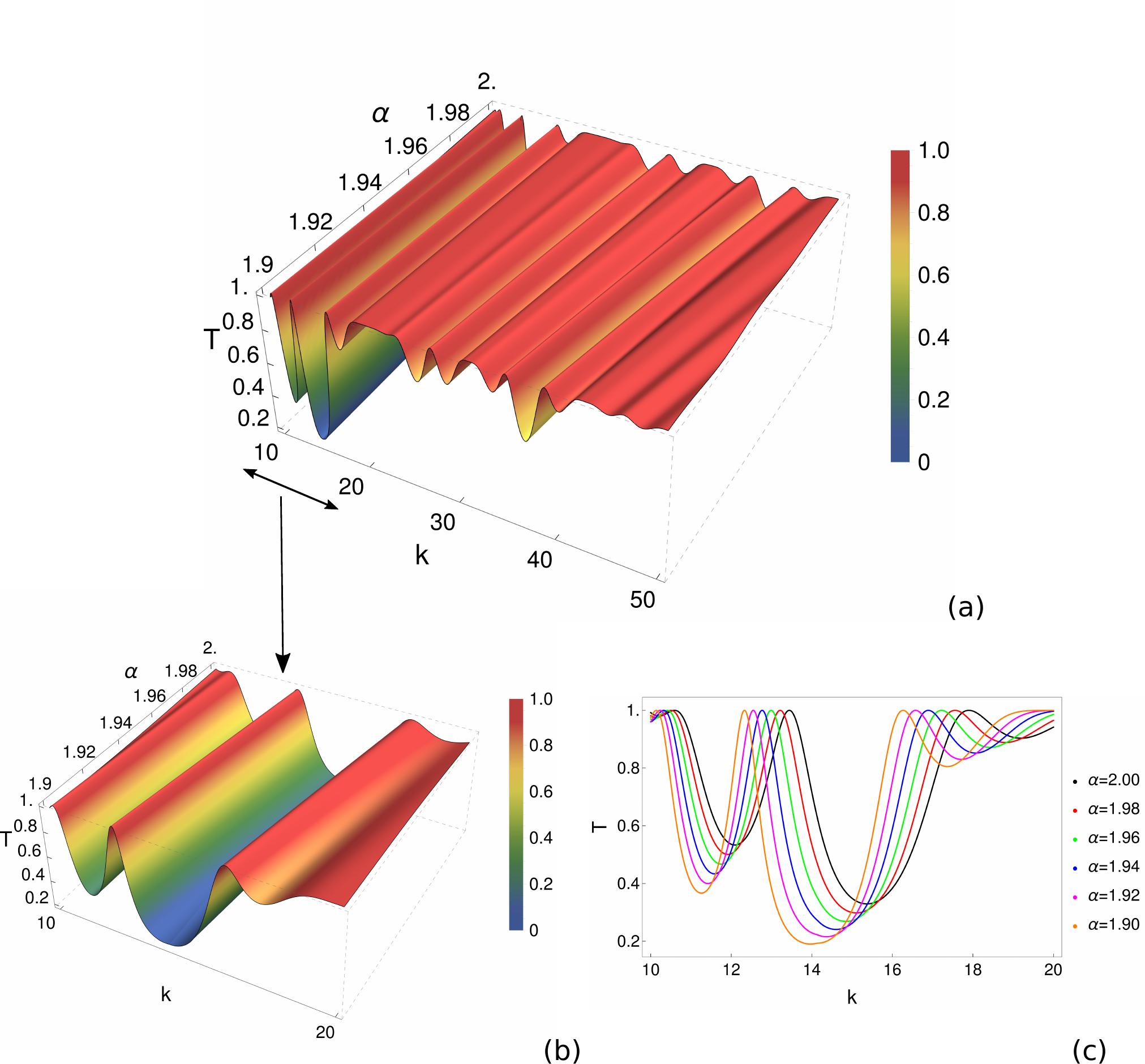}
    \caption{\textit{Plot showing the transmission profile for general Cantor of stage $G=3$, $\rho=3$, $L=1$ and $V=100$ for different value of $\alpha$. It is evident from the plots that as $\alpha$ reduces, the transmission peaks shifts to lower values of wave numbers. (Right Image) 3D plot showing the variation of $T$ with $\alpha$ and $k$ which again depict that the transmission peak occurs at lower $k$ values with reducing $\alpha$.}}
    \label{gc_1_3d}
\end{figure}

\begin{figure}[H]
	\begin{center}
		\centering
		\includegraphics[scale=0.48]{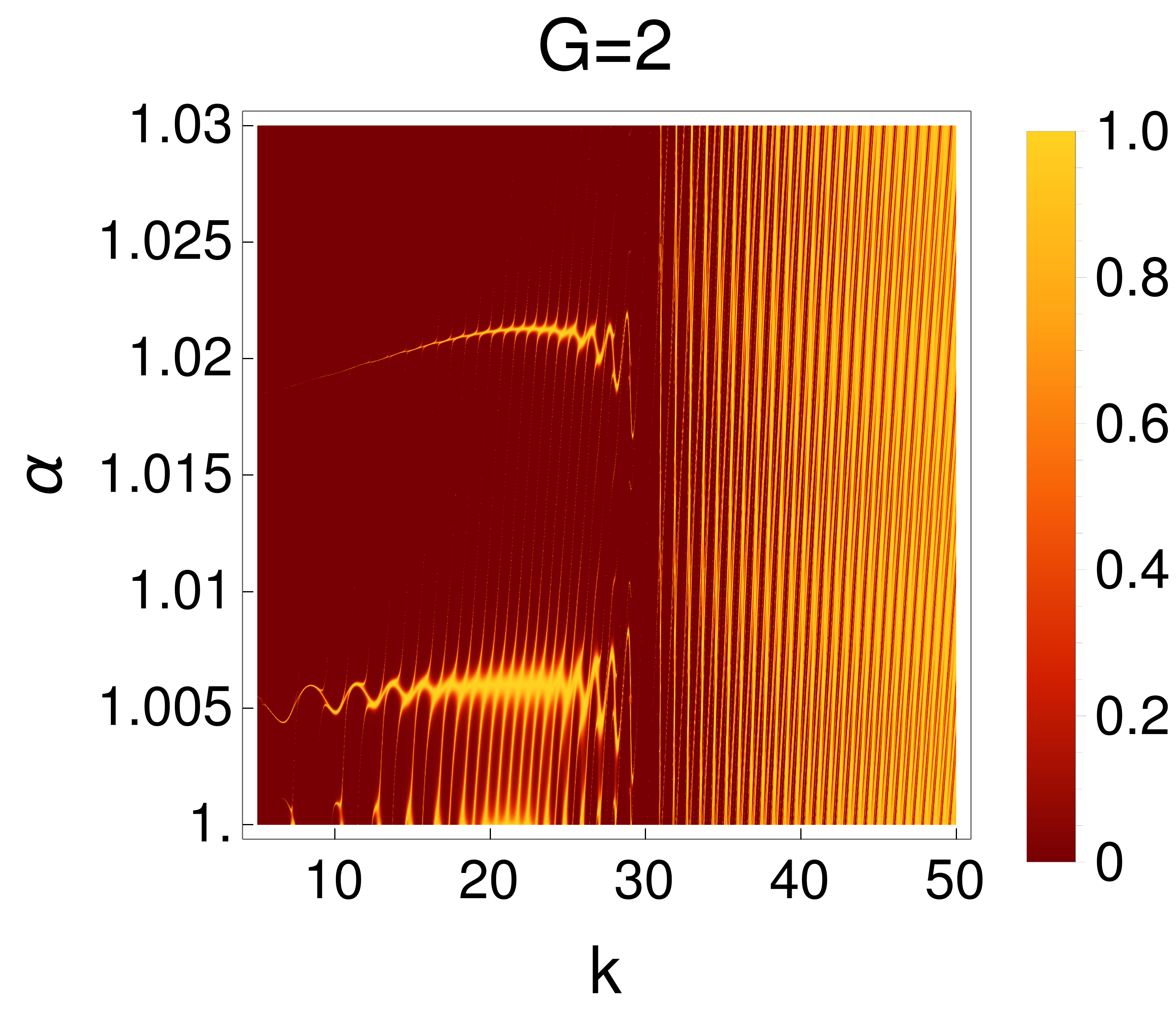} (a)
		\includegraphics[scale=0.48]{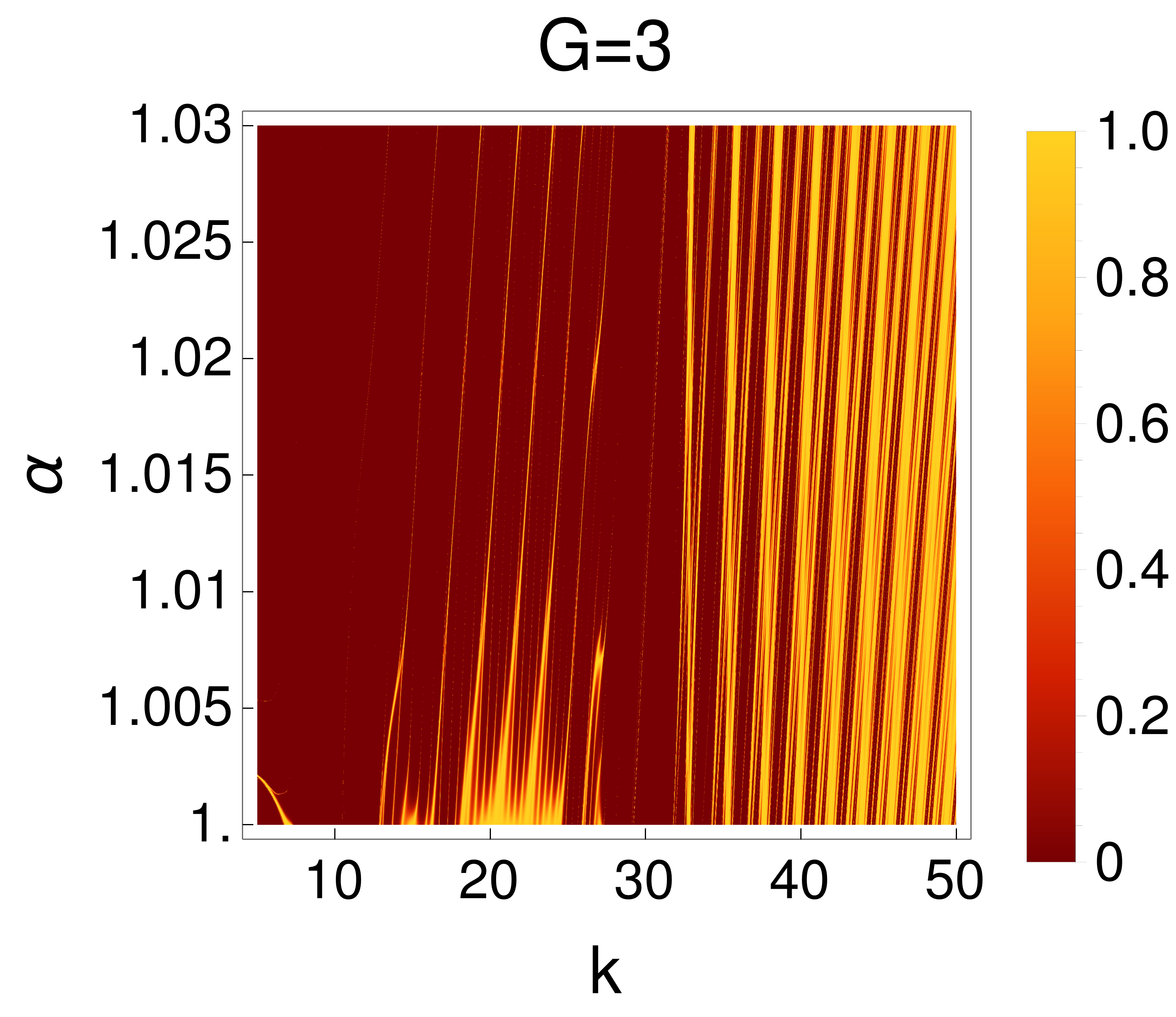} (b)
		\caption{\textit{Density plot showing the variation of transmission amplitude $T$ in $\alpha -k$ plane for $\alpha$ close to 1 for GC potential of different stages $G$. Here  $V=450$, $L=1$ and $\rho=3$. Along with the presence of very sharp transmission peaks, the plots also shows the region of deep valleys in $\alpha -k$ plane where transmission amplitude continuously vanishes. These deep valleys in the transmission profile are the precursor of energy band structure.}}
		\label{GC_DP_alpha_1}
	\end{center}    
\end{figure}

\begin{figure}[H]
    \centering
    \includegraphics[scale=0.72]{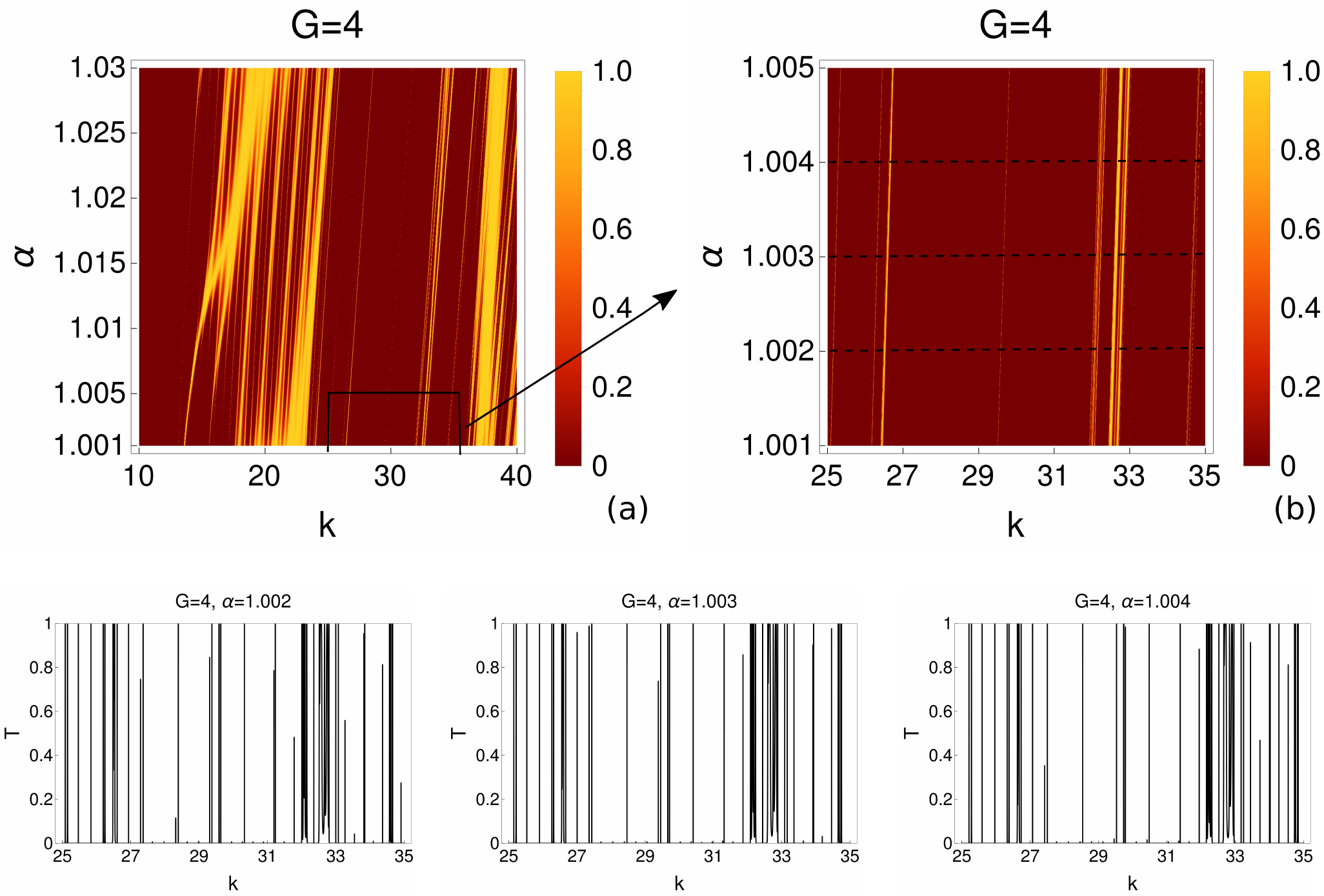}    
    \caption{\textit{(a) Density plot for GC potential of stage $G=4$ showing sharp transmission in $\alpha - k$ plane and energy band features when $\alpha$ is close to 1 and (b) more closer view of density plot for $\alpha$ ranging from $1.001$ to $1.005$. 2D plots illustrate clearly sharp valley for $\alpha=1.002$, $1.003$, and $1.004$. Here potential parameters are $V=450$, $L=1$ and $\rho=3$.}}
    \label{gc_4plot_around}
\end{figure}

In the previous section, we provided some general features of scattering such as emergence of energy bands, increase in the sharpness of transmission resonances with reducing $\alpha$, extreme features of transmission for $\alpha$ near to $1$ etc. for GSVC potential, In this section, we show that such features also exists for GC potential in SFQM. In Fig. \ref{gc_1_3d}, we plot $T(\alpha,k)$ with $\alpha$ and $k$ with potential parameters as $G=3$, $V=100$, $\rho=3$ and $L=1$. A closer look of this figure is shown in Fig. \ref{gc_1_3d}-b for a smaller range of $k$. Similar to the case of GSVC for $\alpha$ near $2$, it is seen that the locus of transmission resonances has a positive slope with increasing $\alpha$. This indicates that the transmission peaks are red-shifted with decreasing values of $\alpha$ for GC potential in SFQM. An exception to this could occur when $\alpha$ is in the vicinity of $1$ (see Fig. \ref{GC_DP_alpha_1}). Fig. \ref{gc_1_3d}-c shows the $2$D plot depicting the variation of $T$ for different $\alpha$ with the same range of $k$ as shown in Fig. \ref{gc_1_3d}-b. Again, similar to the case of GSVC potential, it is observed that the transmission resonances become sharper at lower values of $\alpha$. Fig. \ref{GC_DP_alpha_1} also shows extremely sharp loci of transmission resonances as thin streaks of yellow lines. 
These lines are also separated by deep valleys in $T(k)$ profile which depict the presence of energy band-like features. The deep valleys in $T(k)$ profiles are also shown graphically in Fig. \ref{gc_4plot_around} through the density of $T(\alpha, k)$ in $\alpha-k$ plane as well as $2$D plots for discrete values of $\alpha$. It is to be noted from Fig. \ref{gc_4plot_around}, that many sharp features of transmission resonances that are not visualized due to graphic limitations of capturing very thin streaks of lines are clearly seen in $2$D plots. 
\subsection{Scaling behavior}
\begin{figure}[H]
	\begin{center}
		\includegraphics[scale=0.65]{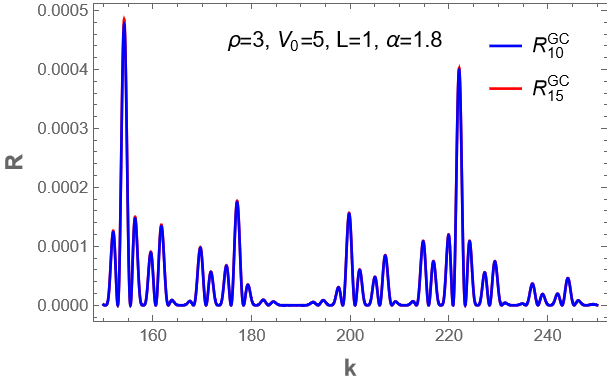} (a) 
		\includegraphics[scale=0.65]{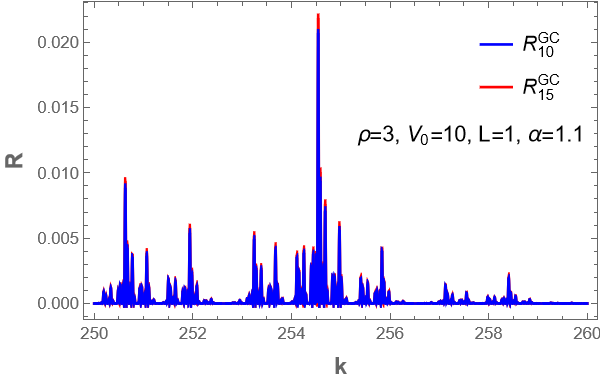} (b)  
		\caption{\textit{Plots showing the reflection amplitudes for GC potential for $G=10$ and $15$ for different $\alpha$ value. The potential height $V_{G}$ is determined from Eq. \ref{vg_values} for GC case. Other potential parameters are shown in the figure. The difference between the two plots is nearly invisible. This shows the convergence of product term of Eq. \ref{vg_zetag} for GC potential.}}
		\label{rg_gc_saturation}
	\end{center}
\end{figure}

This section presents the scaling behavior of the reflection amplitudes $R= \vert r \vert ^{2}$ with $k$ for both types of Cantor potentials considered in the paper. This section also present on how the reflection amplitude behaves when height of the potential $V$ varies in specific manner at each stage $G$. For larger $k$, reflection amplitude $R$ is very small. In this limit, $R$ can be approximated as
\begin{equation}
    R \sim 4^{G} \vert  M_{12} \vert ^{2} \prod_{i=1}^{G} \zeta_{i}^{2}.
    \label{r_small_value}
\end{equation}
Again for larger $k$ we have $\frac{V}{k^2} < < 1$ and upon Taylor expanding, it can be shown in the first order that
\begin{equation}
    \vert M_{12} \vert ^{2} \sim \left ( \frac{\alpha -1}{\alpha} V l_{G} \right )^{2} \frac{1}{k ^{\frac{4(\alpha -1)}{\alpha}}}. 
    \label{m12_small_value}
\end{equation}
Therefore, the expression for $R$ becomes,
\begin{equation}
    R \sim 4^{G} \left ( \frac{\alpha -1}{\alpha} V l_{G} \right )^{2} \frac{1}{k ^{\frac{4(\alpha -1)}{\alpha}}} \prod_{i=1}^{G} \zeta_{i}^{2}.
    \label{r_small_value}
\end{equation}
If $V_{G}$ is the height of the potential at each stage $G$, then it can be shown that the following value of $V_{G}$ keeps the total area of potential barrier (sum of the area of all potential segment at stage $G$) as constant
\begin{equation}
    V_{G} =\frac{L}{2^{G} l_{G}} V_{0},
\end{equation}
where $V_{0}$ is the height of the potential barrier at $G=0$. Substituting the value of $l_{G}$ for GC and GSVC potentials, $V_{G}$ is given by
\begin{equation}
    V_{G}= \left ( \frac{\rho}{\rho -1}\right )^{G} V_{0}, \ \mbox{for GC and}, \  V_{G}= \frac{V_{0}}{q \left ( \frac{1}{\rho}; \frac{1}{\rho} \right)_{G}}   \ \mbox{for GSVC}.
    \label{vg_values}
\end{equation}
If $R_{G}$ is the reflection amplitude at each stage $G$ with potential height of each segment as $V_{G}$ then, it can be shown that (valid for large $k$)
\begin{equation}
    \frac{R_{G}}{L^{2} V_{0}^{2}} \sim  \left ( \frac{\alpha -1}{\alpha}  \right )^{2} \frac{1}{k ^{\frac{4(\alpha -1)}{\alpha}}} \prod_{i=1}^{G} \zeta_{i}^{2}.
    \label{vg_zetag}
\end{equation}
\begin{figure}[H]
\begin{center}
\includegraphics[scale=0.70]{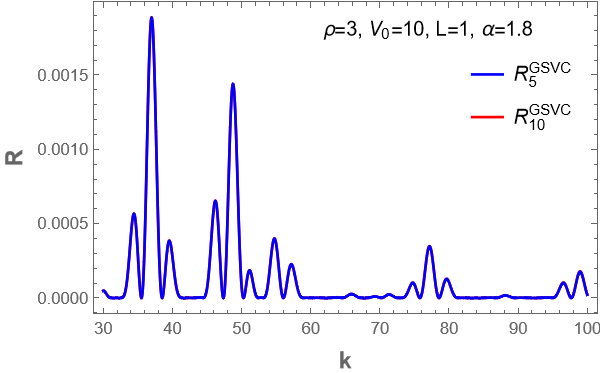} 
\caption{\textit{Plot showing the reflection amplitudes for GSVC potential for $G=5$ and $10$. The potential height $V_{G}$ is determined from Eq. \ref{vg_values} for GSVC potential. Other potential parameters are shown in the figure. The difference between the two plots is invisible. This shows the convergence of the product term of Eq. \ref{vg_zetag}.  }}
\label{rg_gsvc_saturation}
\end{center}
\end{figure}
In Fig. \ref{rg_gsvc_saturation} we show the behavior of $R_{5}^{GSVC}$ and $R_{10}^{GSVC}$. The difference between the two plots is invisible which shows the fast convergence of product term of Eq. \ref{vg_zetag} with increasing $G$ for GSVC case in SFQM. Similar plots are shown for GC case for different $\alpha$ values in Fig. \ref{rg_gc_saturation}. The result shows that the convergence of the product term occurs for GC  case in SFQM with increasing stage $G$. 

\begin{figure}[H]
\begin{center}
\includegraphics[scale=0.45]{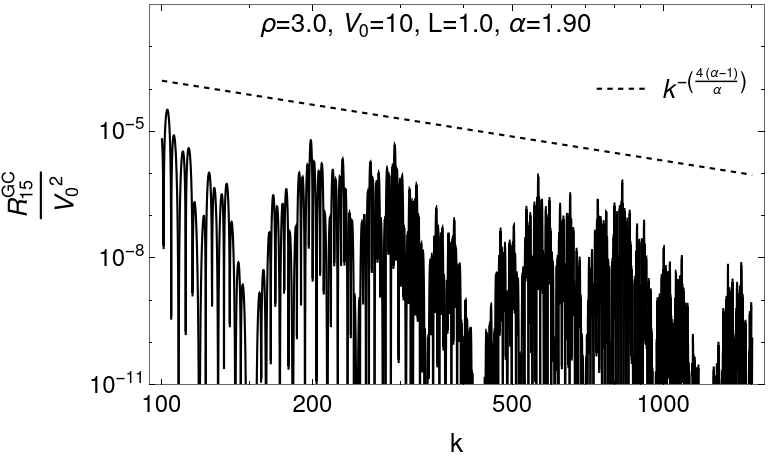} (a)
\includegraphics[scale=0.45]{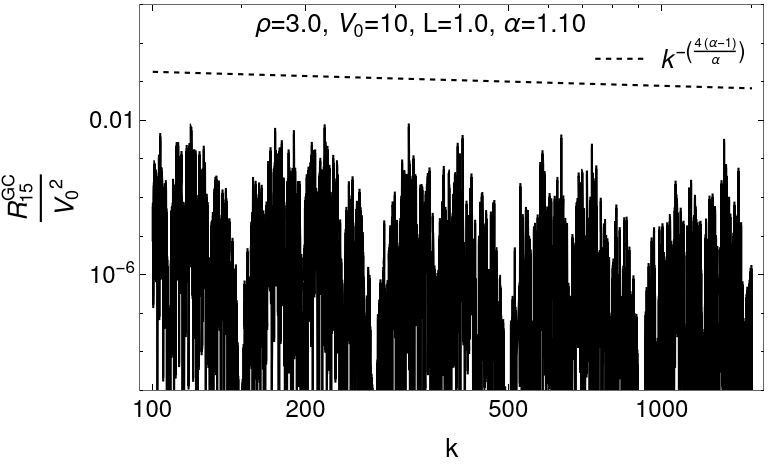} (b)
\includegraphics[scale=0.45]{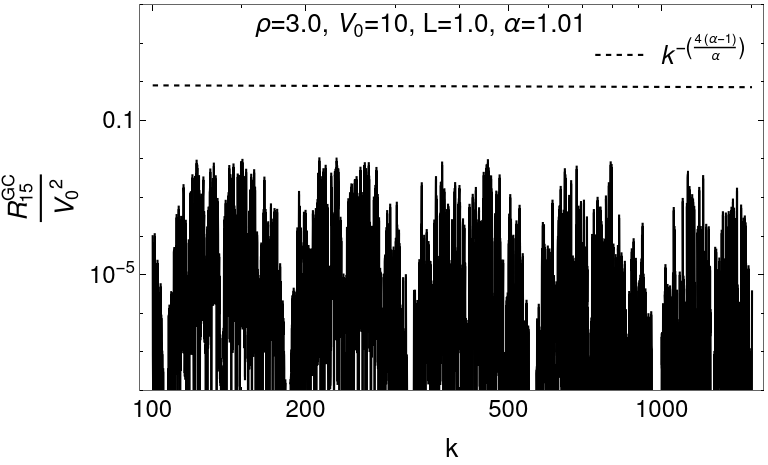} (c)
\caption{\textit{ $\log-\log$ Plots showing the scaling behavior of reflection amplitudes $\frac{R_{G}} {V_{0}^{2}}$ for large $k$ in SFQM in case of general Cantor potential. The dotted curve represent $\frac{1}{k ^{\frac{4(\alpha -1)}{\alpha}}}$.It is observed that at large $k$, $R_{G}$ falls of according to this expression. The potential parameters are shown in the figures.}}
\label{scaling_gc_alpha}
\end{center}
\end{figure}

\begin{figure}[H]
\begin{center}
\includegraphics[scale=0.45]{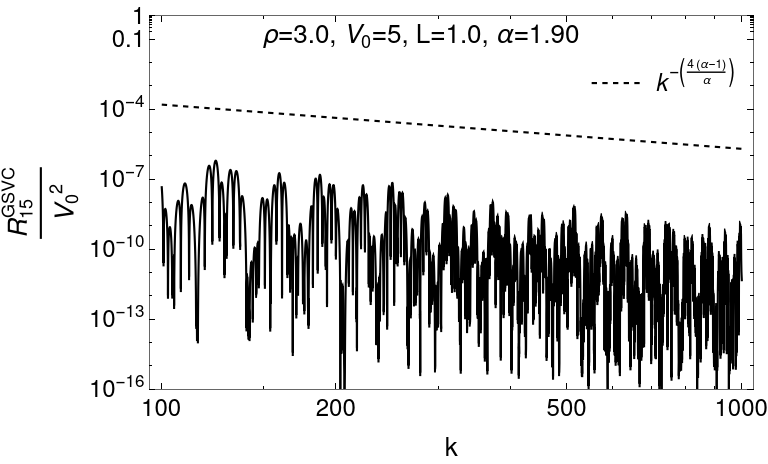} (a)
\includegraphics[scale=0.45]{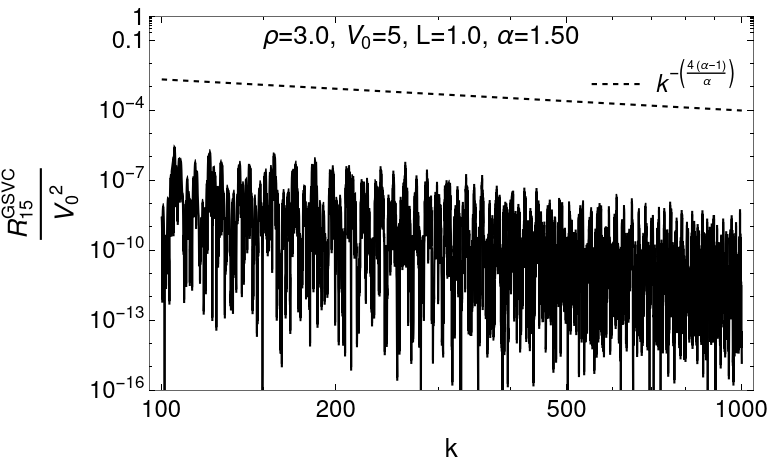} (b)
\caption{\textit{Plots shows reflection amplitude $\frac{R_{G}} {V_{0}^{2}}$ for large $k$ in SFQM ($\alpha=1.90$ and $\alpha=1.50$) for general SVC potential. The dotted curve represent $\frac{1}{k ^{\frac{4(\alpha -1)}{\alpha}}}$.It is observed that at large $k$, $R_{G}$ falls of according to this expression. The potential parameters are shown in the figures.}}
\label{scaling_gsvc_alpha}    
\end{center}
\end{figure}

For Cantor potential in standard QM, this has already shown in earlier work \cite{cantor_f7}. Again, due to the convergence nature of the product term (provided it is evaluated at $V_{G}$) with increasing $G$, it is evident from Eq. \ref{vg_zetag} that $R_{G}$ would scale as $\frac{1}{k ^{\frac{4(\alpha -1)}{\alpha}}}$ for large $G$ and $k$ values. For $\alpha=2$, $R_{G}$ will scale as $\frac{1}{k^{2}}$ which is proven result for standard Cantor potential in standard QM \cite{cantor_f7}. The scaling behavior of $R_{G}$ with $k$ in SFQM is shown graphically for GC potential  in Fig. \ref{scaling_gc_alpha}. An interesting region of interest is when $\alpha$ is near to 1. In this case $R_{G}$ would scale horizontally with $k$ which is indeed the case as shown in Fig. \ref{scaling_gc_alpha}-c. The scaling behavior of $R_{G}$ with $k$ is shown in Fig. \ref{scaling_gsvc_alpha} for GSVC potential for different $\alpha$ values.

\section{Results and Discussions}
\label{R&D}
Fractional quantum mechanics is a fast-developing domain with several applications. We have studied the tunneling features from fractal (general Cantor) and non-fractal (general SVC) potential in space fractional quantum mechanics (SFQM). To the best of our knowledge, this is the first time that the quantum tunneling from fractal potentials in the domain of SFQM are studied. We have considered the generalized form of two kinds of potentials of Cantor family, namely general Cantor (GC) and general Smith-Volterra-Cantor (GSVC) potential. For both the kind of potentials, we have provided close form expressions of transmission amplitudes in SFQM. These close form expressions are expected to provide better understanding of various scattering features in the domain of SFQM from fractal potentials. It is to be noted that the derived expressions are of general type and valid for any potentials of GC and GSVC type in which the `unit cell' is not a rectangular barrier. As long as the transfer matrix of `unit cell' potential is known, the derived expressions can be used to obtain the tunneling amplitudes from such kind of Cantor family potentials. 

In the present study, we have found several new features of scattering, and are reported graphically. The most striking feature is the appearance of energy band structures from fractal potential in SFQM which are absent in the case of standard Cantor fractal and standard SVC potentials in standard QM. Standard fractal potentials based on $\frac{1}{3}$ division of the real segments don't show energy band-like features in standard QM. More surprisingly we noted that a double barrier potential system display energy bands in SFQM. We have reported the emergence of band structures for both the type, GC and GSVC potentials. However, it is to be noted that these band like features appear in the extreme range of Levy index $\alpha$ close to the vicinity of $1$. In this range of $\alpha$, extremely sharp transmission resonances are found to occur for both type of potentials.   

Fractal potentials are known to display sharp transmission resonances, This feature is further amplified in the domain of SFQM. It is found that the sharpness of the transmission resonances further increases with a decrease in Levy index $\alpha$. In comparison, GSVC potential displays more sharp transmission resonances as compared to GC potential in standard QM as well as in SFQM. Also for the case of GSVC potential, it is observed that the profile of transmission amplitudes saturates with increasing stage $G$.  The reason for this is due to the fact that a consecutively smaller fraction of the remaining previous segments is removed at each stage G for GSVC potentials Therefore for higher $G$, only very thin portions are removed from previous segments as compared to the case when $G$ is small. This leads to the saturation of the tunneling profile with $k$ for higher $G$. However, this behavior is found to be different near $\alpha=1$. For $\alpha \sim 1^{+}$, the tunneling profile saturates only for $E>V$.   

Another interesting feature is the scaling behavior of reflection amplitude. We have shown analytically that for large $k$, the reflection coefficient scale as $\frac{1}{k^{\frac{4(\alpha-1)}{\alpha}}}$ for GC and GSVC potential provided that total area of the potential regions remain constant at different stages $G$. Also for such case, the reflection amplitude converges as $G$ increases for both GC and GSVC systems. 

{\it \bf{Acknowledgements}}:\\ 
  The present investigation has been carried out under financial support from BHU RET fellowships to VNS from Banaras Hindu University (BHU), Varanasi. BPM acknowledges the support from the Research grant under IoE scheme (Number- 6031), UGC-Govt. of India. MH acknowledges supports from SPO-ISRO HQ for the encouragement of research activities.

\newpage
\bibliographystyle{elsarticle-num}
\bibliography{References}

\end{document}